\documentclass[12pt]{iopart}
\usepackage[utf8]{inputenc}
\usepackage[sorting=none]{biblatex}

\expandafter\let\csname equation*\endcsname\relax

\expandafter\let\csname endequation*\endcsname\relax
\usepackage{amsmath}
\usepackage{amssymb}
\usepackage{mathtools}

\usepackage{hyperref}
\usepackage{graphicx}

\newcommand{\avg}[1]{\left\langle #1 \right\rangle}

\begin{document}

\title{Reduced model for H-mode sustainment in~unfavorable $\mathbf{ \nabla B}$ drift configuration in~ASDEX Upgrade}
\author{O. Grover$^1$ , T. Eich$^{2,1}$ , P. Manz$^{3,1}$ , W. Zholobenko$^1$ , T. Happel$^1$, T. Body$^2$, U. Plank$^1$, P. Ulbl$^1$,   ASDEX Upgrade team$^4$ }
\address{$^{1}$ Max Planck Institute for Plasma Physics, Boltzmannstr. 2, 85748 Garching, Germany}
\address{$^2$ Commonwealth Fusion Systems, Devens, MA, USA}
\address{$^3$ Institute of Physics, University of Greifswald, Felix-Hausdorff-Str.~6, 17489 Greifswald, Germany}
\address{$^4$ See author list of \textit{U. Stroth et al. 2022 Nucl. Fusion \textbf{62} 042006}}
\ead{ondrej.grover@ipp.mpg.de}

\begin{abstract}
    
A recently developed reduced model of H-mode sustainment based on interchange-drift-Alfv\'en turbulence description in the vicinity of the separatrix matching experimental observations in ASDEX Upgrade has been extended to experiments with the unfavorable $\nabla B$ drift. The combination with the theory of the magnetic-shear-induced Reynolds stress offers a possibility to quantitatively explain the phenomena.

The extension of the Reynolds stress estimate in the reduced model via the magnetic shear contribution is able to reproduce the strong asymmetry in the access conditions depending on the ion  $\nabla B$ drift orientation in agreement with experimental observations.
The Reynolds stress profile asymmetry predicted by the magnetic shear model is further extended by comparison with GRILLIX and GENE-X simulations matched with comparable experiments in realistic X-point geometry. The predictions of the radial electric field well depth and its difference between the favorable and unfavorable configurations at the same heating power from the extended model also show consistency with experimental measurements.
\end{abstract}


\submitto{\NF}

\maketitle

\section{Introduction}

Access to the high confinement regime (H-mode)~\cite{ASDEX-H-mode} in tokamaks is today considered one of the key ingredients for developing a fusion reactor. However, to this day several aspects of the physics of accessing H-mode are still not fully understood.

One of them is the dependence on the X-point position relative to the main plasma, known for more than 30 years since the times of ASDEX~\cite{ASDEX-H-mode} and reproduced in many other devices such as ASDEX Upgrade~\cite{Ryter_1998}, Alcator C-Mod~\cite{Greenwald_1997} and DIII-D~\cite{carlstrom1997h}. In the so-called favorable $\nabla B$ drift configuration associated with the ion $\nabla B$ drift pointing \textit{towards} the active X-point, ample empirical evidence has led to the development of scaling laws for the threshold power necessary to access the H-mode.

However, in the unfavorable configuration (ion $\nabla B$ drift pointing \textit{away} from the active X-point) the power threshold is typically observed to be at least 2 times larger than in the favorable configuration~\cite{Snipes_1996,McKee_2009,Cziegler2017PRL}.   Furthermore, the  radial electric field $E_r$ shear at the plasma edge has been observed to be significantly weaker in  the unfavorable configuration with respect to the favorable configuration at the same heating power and density, see~\cite{Plank_Er_2023} and references therein..

Understanding this asymmetry in H-mode access with respect to the X-point orientation is significant both for testing models of H-mode access as well as assessing the operational range of alternative confinement regimes such as the I-mode, which is typically accessed in the unfavorable configuration due to the  H-mode being less accessible~\cite{Hubbard_2016}.

Attempts at explaining the behavior often focus on the experimental observation that the radial electric field $E_r$ profile at the edge of the plasma before the transition to H-mode is very different between the two configurations.
Several theories and physical mechanisms for explaining this behavior have been proposed, among others mechanisms dependent on drift directions such as SOL flows setting the boundary conditions for $E_r$~\cite{LaBombard2005SOLflows}, neutral penetration~\cite{CHANKINA2017273}, or mechanisms based on the X-point-induced asymmetry such as ion orbit losses~\cite{Battaglia_2013}, and acceleration due to  different Reynolds stresses~\cite{Fedorczak_2012}. The magnetic-geometry-induced Reynolds stress theory was able to recover the difference in $E_r$ profiles in Tore Supra~\cite{Fedorczak_2013}, and more recently an extended version of the model similarly recovered $E_r$ profiles in WEST~\cite{Peret2022}.

In this article a somewhat different approach based on the separatrix operational space (SepOS) model~\cite{Eich_2021} is taken, rather focusing on an established H-mode with a typical  $E_r\sim \nabla p_i /(e n_i)$ profile~\cite{Viezzer_2014}, but with  turbulence suppression rates  facilitated by the Reynolds stress~\cite{Manz2012PoP-suppr-model}. The Reynolds stress in turn is assumed to depend on the X-point configuration along the lines of Reynolds stress symmetry breaking as in~\cite{Fedorczak_2012} and~\cite{StaeblerPoP2011}. Therefore, this article focuses conceptually rather on H-mode sustainment and thereby the H-L back-transition, rather than on the H-mode access and the L-H transition. The possible impact of any hysteresis effects will be discussed later.

The article is structured as follows: In~\sref{sec:sepos-empirical} the empirical observations in ASDEX Upgrade are interpreted in the framework of the SepOS model and differences between the favorable and unfavorable configurations are empirically characterized by a single factor. In order to find a possible explanation for this factor, the Reynolds stress X-point dependence mechanism from~\cite{Fedorczak_2012} is briefly reviewed and recast in a more intuitive form in order be included in the SepOS in~\sref{sec:coupling-shear-RS}. To quantify and better understand the Reynolds asymmetry, state-of-the-art full-f simulations are analyzed in \sref{sec:equilibrium-RS-model}, extending the preliminary analysis in~\cite{Manz2018}. Finally, the insight gained from simulations is combined with the SepOS model to consistently explain the empirical factor. As a consistency check, a proof-of-principle reconstruction of differing L-mode $E_r$ profiles from~\cite{Plank_Er_2023} is attempted in~\sref{sec:Er-L-mode-consistency}.

The reader may notice that the article is long and dense with equations, with many  plus and minus signs. We beg for the reader's understanding and patience, since surely one must admit that a solution to a problem eluding understanding for the last 30 years is unlikely to be trivial. 

\section{Empirical $\mathbf{ \nabla B}$ difference from the perspective of the SepOS model}
\label{sec:sepos-empirical}

Conceptually, the H-mode turbulence suppression criterion derived in~\cite{Manz2012PoP-suppr-model} and employed in the SepOS model~\cite{Eich_2021}  is given by the following balance of the energy transfer from turbulence to the shear flow (which suppresses the turbulence) on the left-hand-side and the turbulent energy production on the right-hand-side:

\begin{equation}
    \frac{1}{1+\delta_{\phi,p_e}^2}\langle u_y\rangle \partial_x \langle \tilde u_x\tilde u_y\rangle > \gamma_e (E_{tk} + E_{te}) +\gamma_i E_{ti} 
    \label{eq:H-suppr-conceptual}
\end{equation}
with the Reynolds stress  $\langle \tilde u_x\tilde u_y\rangle $ and the mean shear flow $\langle u_y\rangle$ on the left-hand-side, representing together the so-called Reynolds work - transfer of turbulent energy to the flow. The $x$ and $y$ coordinates represent the locally field-aligned radial and bi-normal (to both the magnetic field and $x$, i.e. close to poloidal) directions, respectively.  The transfer is made less effective by low adiabaticity represented by the cross-phase $\delta_{\phi,p_e}$ between the potential and pressure. The turbulent energy production is approximated by the characteristic quasi-linear growth rates  of electron $\gamma_e$ and ion $\gamma_i$ modes and the kinetic $E_{tk}$  and free energies $E_{te},E_{ti}$. 

Through further quasi-linear approximations and by normalizing the equations to the kinetic energy scale $E_{tk}$ the criterion in normalized units can be understood as a relation between an effective turbulence energy suppression rate $\gamma_\mathrm{suppr,eff}$ (stabilizing term) and an effective turbulence energy production rate $\gamma_\mathrm{prod,eff}$ (destabilizing term)
\begin{align}
    \gamma_\mathrm{suppr,eff} &> \gamma_\mathrm{prod,eff}    \label{eq:H-suppr-gamma-eff}\\
    \frac{k_\mathrm{ES}\tau_i\Lambda_{pi}}{1+\left(\frac{\alpha_t}{\alpha_c} k_\mathrm{ES}\right)^2} &> \frac{\alpha_t}{\alpha_c} \left(k_\mathrm{ES}^2+\frac{1}{2}\right)+\frac{1}{2k_\mathrm{ES}^2}\sqrt{\omega_B\tau_i\Lambda_{pi}}\notag\\
    \alpha_t =& 2.22\cdot10^{-18} R q_{cyl}^2 \frac{n_e}{T_e^2} (1+\tau_i) Z_\mathrm{eff}\propto q_{cyl} \nu^*\notag\\
    k_\mathrm{ES}^2 =& \frac{m_i}{m_e}\frac{2\mu_0 n_e T_e}{B^2}\propto\beta\notag\\
    \tau_i=&\frac{p_i}{p_e}=\frac{T_i}{ZT_e};\quad\tau_i\Lambda_{pi}=\frac{\nabla p_i}{\nabla p_e}\approx 1\notag\\
    \alpha_c=&\kappa^{1.2}(1+1.5\delta)\notag\\
        \omega_B =& 2 \frac{\lambda_{p_e}}{R}\notag
\end{align}
with the normalized collisionality turbulence parameter $\alpha_t$ (closely related to the cylindrical safety factor $q_{cyl}$ and the typical edge electron collisionality $\nu^*\propto q_{cyl}R n_e/T_e^2$)  describing the cross-phase of potential and pressure perturbations (i.e. drift-wave or interchange transport), the typical normalized electrostatic wavenumber scale $k_\mathrm{ES}^2\propto \beta$ (previously called $k_\mathrm{EM}$\footnote{the original name in~\cite{Eich_2021} comes from the derivation of a typical wavenumber scale below which electro-magnetic (EM) effects become significant. However, subsequently a factor 2 larger wavenumber scale is used for the quasilinar approximation of electro-static (ES) turbulence behavior (i.e. 2 times larger wavenumbers are expected to be far from EM effects, hence should be dominantly ES), therefore, the EM subscript is somewhat misleading and is replaced with the more accurate ES}), the critical normalized pressure gradient $\alpha_c$ capturing shaping dependencies (of elongation $\kappa$ and triangularity $\delta$) and the normalized  curvature  $\omega_B$, set by the major radius $R$ normalized by the electron pressure gradient length $\lambda_{p_e}$,  describing interchange growth. The ion to electron pressure gradients ratio $\tau_i\Lambda_{pi}$ (assumed to be close to 1) represents the mean flow approximated by the diamagnetic flow $\langle u_y\rangle\approx\nabla p_i/neB$ (which normalizes to $\tau_i\Lambda_{pi}\approx 1$) on the left-hand-side and approximates an ITG-like growth rate on the right-hand-side. The approximation of the growth rate by $\propto\sqrt{\omega_B}$ is consistent with curvature in the vicinity of the outer midplane (OMP). For details the reader is referred to Ref.~\cite{Eich_2021}.

Note that the \textit{effective} suppression and production rates are by construction rather close to a shearing rate $\partial_x u_y$ and linear growth rates, respectively, reminiscent of the more commonly applied shear suppression criteria $\partial_x u_y \sim \gamma$~\cite{Waltz94shear}. However, here these effective rates are not exactly the same thing due to being constructed from energy transfer arguments. 

For the present discussion it suffices to point out that all the terms in \eqref{eq:H-suppr-gamma-eff} are functions of measurable quantities: Firstly the equilibrium parameters such as the cylindrical safety factor $q_{cyl}$, major radius $R$ and shaping (average elongation and triangularity), and secondly of kinetic quantities: the electron temperature $T_e$ and density $n_e$ and  the edge pressure gradient length $\lambda_{p_e}:=-p_e/\nabla_r p_e$. The three kinetic quantities  are  estimated at the very edge of the confined plasma in the vicinity of the separatrix in ASDEX Upgrade from Spitzer-H\"arm power balance analysis and Thomson Scattering profiles for stationary plasma discharge windows of $\sim 200-400$~ms  as explained in detail in Ref.~\cite{Eich_2021}. 

\begin{figure}[htb]
    \centering
    \includegraphics[width=\linewidth,height=0.7\textheight,keepaspectratio]{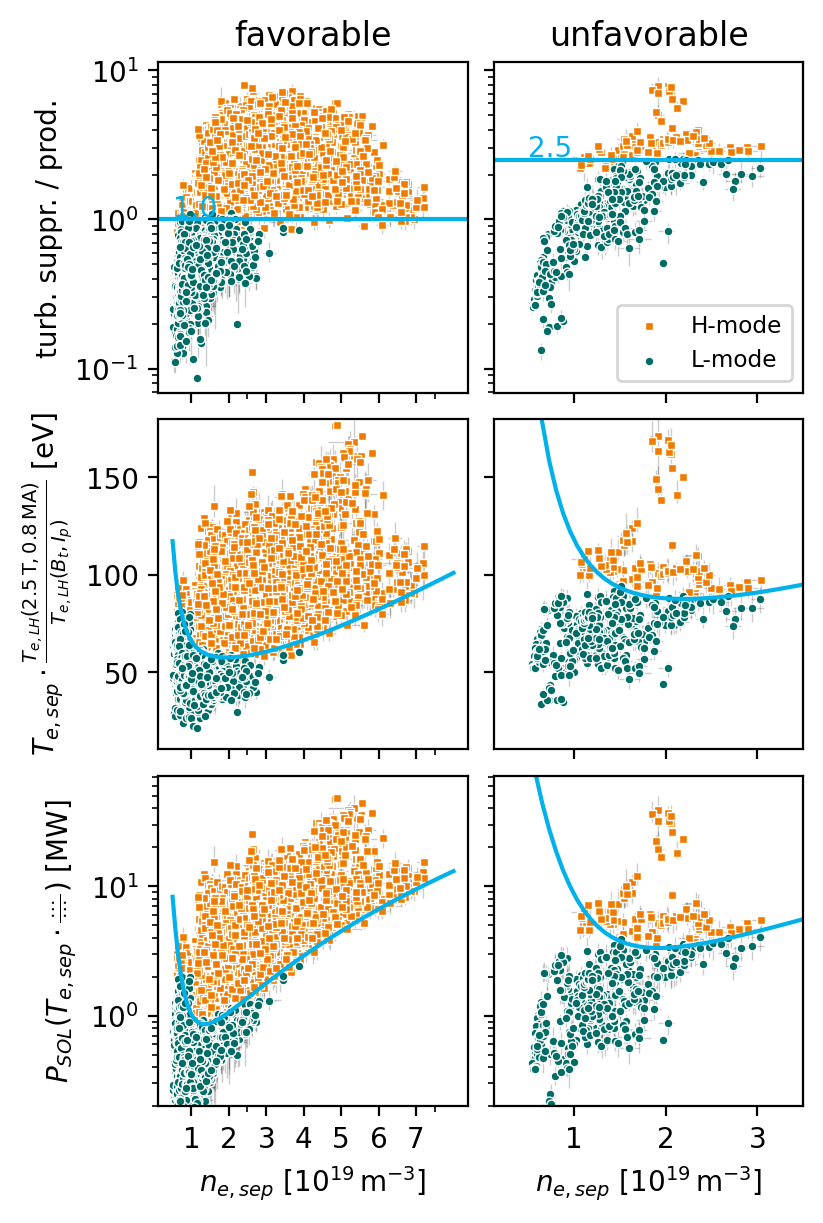}
    \caption{H-mode suppression criterion compared with the AUG separatrix quantities database for the favorable and unfavorable in the left and right columns, respectively. The  first row shows the ratio of the two sides of \eqref{eq:H-suppr-gamma-eff}, i.e. the $\gamma_\mathrm{suppr,eff} / \gamma_\mathrm{prod,eff}$ ratio. The second row shows the database points in the $(n_e,T_e)$ space normalized to a standard scenario and the blue H-mode suppression line corresponding to \eqref{eq:H-suppr-gamma-eff} for the standard scenario. The third row shows the corresponding power crossing the separatrix $P_{SOL}$ estimated from the normalized temperatures. Note that the normalization for some discharges (lower toroidal field scenarios) results in a $P_\mathrm{SOL}$ not actually achievable in ASDEX Upgrade.}
    \label{fig:H-mode-suppr-fav-unfav}
\end{figure}

For each of the windows where this estimation is done  the ratio of $\gamma_\mathrm{suppr,eff} / \gamma_\mathrm{prod,eff}$ as the criterion for \eqref{eq:H-suppr-gamma-eff} for H-mode turbulence suppression can be thus evaluated and compared with the actual identified confinement regime. This is shown in the first row of~\fref{fig:H-mode-suppr-fav-unfav} for several combinations of magnetic fields and plasma currents and for the favorable and unfavorable drift cases. The range of these parameters of the databases is summarized in~\tref{tab:sepos-db}. All the discharges considered here are in deuterium, typically shortly after boronization with no additional impurity seeding, hence a low effective charge $Z_\mathrm{eff}\approx 1.24\pm 0.13$ (part of the displayed errorbars) is assumed  as in~\cite{Eich_2021}. While $Z_\mathrm{eff}$ is not actually known at the separatrix location, the comparison with a core $Z_\mathrm{eff}$ estimate shows no systematic difference between the favorable and unfavorable configuration datasets presented here and agrees with the assumed  $Z_\mathrm{eff}$ range. For the unfavorable cases both lower and upper single null (LSN and USN) discharges with corresponding unfavorable field orientations are used.

\begin{table}[htb]
    \centering
    \begin{tabular}{c|rrrrrr}
         $\nabla B$ case & points & $B_t$ [T] & $I_p$ [MA] & $q_{cyl}$ \\\hline
         favorable & 18172 & 1.4-3.1 & 0.4-1.2 & 2.7-8.4\\
         unfavorable & 2001 & 1.8-3 & 0.8-1 & 3-6.2
    \end{tabular}
    \caption{Range of global equilibrium parameters of the favorable and unfavorable separatrix operational space databases}
    \label{tab:sepos-db}
\end{table}

The most important observation is that the ratio $\gamma_\mathrm{suppr,eff} / \gamma_\mathrm{prod,eff}$ at the horizontal boundary separating L-modes and H-modes remains constant across a wide density range, as well as a wider range of magnetic fields and currents as summarized in \tref{tab:sepos-db}, effectively spanning the whole operational range of ASDEX Upgrade. For the favorable case the ratio is 1 by construction, since \eqref{eq:H-suppr-gamma-eff} was constructed in such a way so as to describe the data. In other words, this is a data-driven approach which attempts to capture the scaling of the data through physics-informed parameters. But the fact that a constant, albeit higher, ratio is also found for the unfavorable cases is a non-trivial result.  This  shows that not only these normalized variables manage to capture the various (and even non-monotonic) scalings of the L-H boundary with current, magnetic field, density, etc., but additionally this specific ratio of the terms apparently contains the physics missing to explain the difference between the favorable and unfavorable cases.   

The ratio near the L-H boundary for the unfavorable case appears to be about 2.5. This means that for a given effective production rate in the unfavorable case one needs a 2.5 times higher suppression rate compared to an equivalent favorable case for H-mode to exist. This can be also understood as the suppression rate being only $1/2.5=0.4=40\%$ as efficient as in the equivalent favorable case with the same production rate.
The main question now becomes: What does this factor mean and why does it have this value?

Before this question is tackled, it is illustrative to show how this larger factor in the unfavorable case impacts the power threshold.  To this end, the criterion \eqref{eq:H-suppr-gamma-eff} can be interpreted also in the $(n_e, T_e)$ operational space which is related to the power entering the SOL $P_\mathrm{SOL}\propto T_e^{3.5}$.
For a given equilibrium (i.e. fixed $q,R$, shaping) and for each $n_e$ value \eqref{eq:H-suppr-gamma-eff} can be solved as a non-linear equation for $T_e$, given a relation for the gradient length $\lambda_{p_e}(T_e, n_e,\dots)$. For these cases the  H-mode ELMy scaling $\lambda_{p_e}\propto \rho_{s,pol}(1+C\cdot \alpha_t^{1.9})$ is used as in~\cite{Eich_2021} which describes the widening of the gradient lengths at higher collisionalities. The solution at each density $T_\mathrm{e,LH}(n_e)$ for the standard AUG scenario and shape with magnetic field $B_t=2.5$~T and current $I_p=0.8$~MA then becomes the blue line, similar to an L-H power threshold, in the second row of~\fref{fig:H-mode-suppr-fav-unfav}. Since the experimental $T_e$ points come from a wider range of current and magnetic fields and shapes, they are shown normalized by the ratio of $T_\mathrm{e,LH}(n_e)$ for the standard $I_p, B_t$ and their scenario $T_\mathrm{e,LH}(n_e,2.5\,\mathrm{T},0.8\,\mathrm{MA})/T_\mathrm{e,LH}(n_e,B_t,I_p)$. The blue line again well separates L-mode and H-modes. Most importantly, the line is significantly higher (at the same density) in the unfavorable case than in the favorable case. 

Through the formula $P_\mathrm{SOL}\propto T_e^{3.5} \lambda_q(\alpha_t)$ the normalized temperature translates to a significant increase in the L-H power threshold as shown in the third row. Interestingly, due to the widening of the gradient and power widths dependent on $\alpha_t$, the $P_\mathrm{SOL}$ is  about 2.5 times larger at higher densities in the unfavorable configuration, though at lower densities (less widening) it could be even more than 4 times higher compared to the favorable case at the same density. Therefore, it qualitatively reproduces this difference in power threshold increase in the low and high density branches reported in~\cite{Plank_Er_2023}.

One possible explanation for the factor 2.5 is that either the left or right-hand side of \eqref{eq:H-suppr-gamma-eff} is actually missing some factor. For instance, if the stabilizing term on the left-hand side of \eref{eq:H-suppr-gamma-eff} had an additional factor of 0.4 (i.e. was only 40 \% efficient) in the unfavorable and 1 in the favorable cases, then \eref{eq:H-suppr-gamma-eff} could describe both configurations consistently. An obvious candidate for such an explanation is the pressure gradient ratio $\tau_i\Lambda_{pi}$ which was only assumed to be close to 1 in typical discharges in the favorable case.  However, dedicated measurements of this ratio in the vicinity of the separatrix show that it  is very unlikely 
that the ratio would be systematically $\tau_i\Lambda_{pi}\approx 0.4$ only in the unfavorable cases.

Therefore, a different mechanism for changing the terms was sought.

In the next two sections \ref{sec:coupling-shear-RS} and \ref{sec:equilibrium-RS-model} an explanation for the factor 2.5 is sought by extending the  model to have the energy transfer through the Reynolds stress  dependent on the X-point geometry   based on ideas  from~\cite{Fedorczak_2012,StaeblerPoP2011}.
However, the result of these sections is essentially just a single numerical factor  and several general observations, and so the impatient reader may jump to section~\ref{sec:sepos-RS-tilt} to see how these are used in explaining the 40\% factor.

\section{Coupling with the shear-induced Reynolds stress model}
\label{sec:coupling-shear-RS}

 Because the Reynolds stress enters in the stabilizing term of \eqref{eq:H-suppr-conceptual}, it follows that the theory of the impact of the X-point asymmetry on the Reynolds stress as described in Ref.~\cite{Fedorczak_2012} is an obvious candidate for extending the SepOS model. In the following the  pieces of the two theories ~\cite{Eich_2021} and~\cite{Fedorczak_2012,Fedorczak_2013} relevant for the coupling are reviewed and the theories are shown to be ideally suited for such a coupling.

As seen  in~\eqref{eq:H-suppr-conceptual}, the terms are approximated  
mainly through characteristic wavenumbers and linear growth rates under the assumption of $k_x=k_y=k_\perp$ set to $k_\mathrm{ES}$ which as will be illustrated later in~\sref{sec:sepos-RS-tilt} is generally 
consistent with the shear suppression criterion. 
However, for the following considerations based on eddy tilting in~\cite{Fedorczak_2012,Fedorczak_2013} it is necessary to distinguish $k_y$ and $k_x$.

In particular, the Reynolds stress  is approximated in~\cite{Eich_2021,Fedorczak_2013,Fedorczak_2012} through
\begin{equation}
     \langle \tilde u_x\tilde u_y\rangle \approx -  \langle (k_x k_y)\tilde \phi^2 \rangle 
    \label{eq:RS-DALF}
\end{equation}
 because  the turbulent velocity fluctuation amplitudes are approximated by typical scales of fluctuating $E\times B$ velocities $\tilde u_x \approx i k_y\tilde\phi$ and $\tilde u_y\approx -i k_x\tilde\phi$ with the fluctuating plasma potential $\tilde \phi$. The not-so-trivial derivation of the minus sign on the right-hand side of \eqref{eq:RS-DALF} is discussed in more detail in \sref{sec:RS-sign}. The magnitude of the magnetic field $B$ is ``hidden'' by the normalization of wavenumbers to the hybrid (sound) gyroradius $\rho_s=\sqrt{m_iT_e}/eB$, i.e. $k_x$ is in fact $k_x\rho_s$ in SI units. The actual amplitude of the plasma potential fluctuation energy $\tilde \phi^2$  in \eqref{eq:H-suppr-gamma-eff} cancels through the normalization by the kinetic energy, therefore, only the description of the Reynolds stress through wavenumber scales is sought for the coupling. The averaging $\langle\dots\rangle$ brackets represent an ensemble average over turbulent fluctuations and also a flux surface average, because the   Reynolds stress can poloidally vary on a flux surface, as will be shown later, and only the ``net'' suppression effect is of interest.

\begin{figure}
    \centering
    \includegraphics[width=\linewidth]{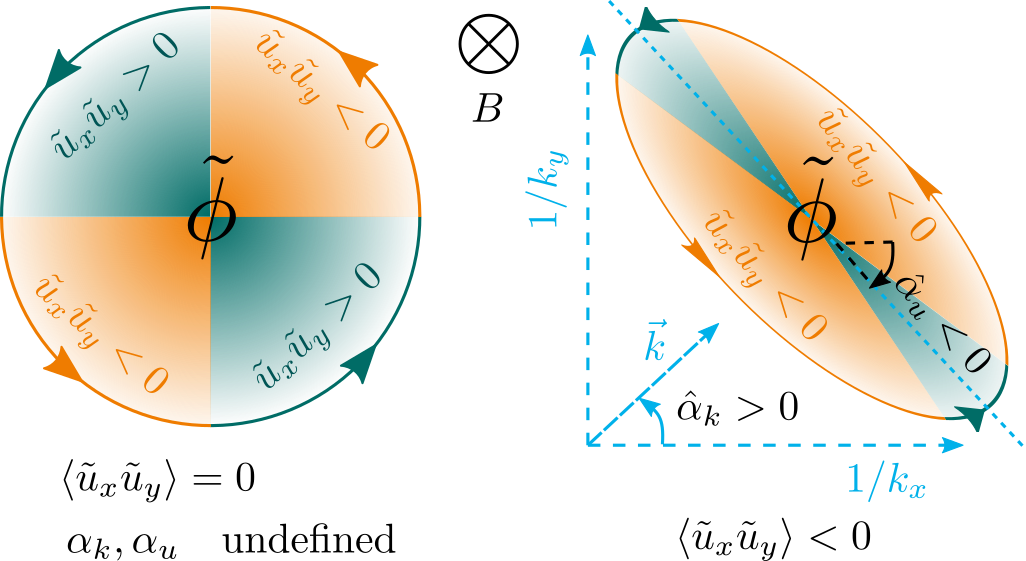}
    \caption{Sketch of how the shape of an $E\times B$ eddy (vortex) around a potential perturbation $\tilde \phi $ determines its contribution to the Reynolds stress $\langle \tilde u_x\tilde u_y\rangle_t$. With no co-linearity of velocities (left), the average covariance of velocities over all its regions is 0, while with co-linearity (right) - specifically anti-correlation - the regions with one $(\tilde u_x\tilde u_y)$ sign dominate. Sketch inspiration from~\cite{ManzHabil2018}. The blue arrows indicate the corresponding wavenumber vector $\vec k$ and wavelength $1/k$ scales in a plane-wave-like approximation. The angles with the hat accent $\hat \alpha$ correspond to the tilt ratios through $\alpha=\tan(\hat\alpha)$}
    \label{fig:eddy-tilt}
\end{figure}

The poloidal profile of the Reynolds stress can be deduced from the relationship with the structure tilt: At a given poloidal location on a flux surface the covariance of the turbulent velocity fluctuations resulting in the Reynolds stress is not only due to the amplitude (variance) of the fluctuations, but also due to the  correlation of the velocity fluctuations. The simplest model for correlated velocity fluctuations is a linear correlation $\tilde u_y \propto \tilde u_x$. Locally (at a given poloidal location), such a co-linearity can be related to the average  tilt of stretched (sheared) structures as illustrated in~\fref{fig:eddy-tilt}. The proportionality constant  $\alpha_u$ can be related to the corresponding wavenumbers through
\begin{eqnarray}
     \alpha_u=\frac{\tilde u_y}{\tilde u_x}\approx\frac{-i k_x\tilde\phi}{i k_y\tilde\phi}=-\frac{k_x}{k_y}=-\alpha_k
    \label{eq:alpha-k}\\
    \tan(\hat\alpha_u) =\alpha_u \qquad\, \tan(\hat\alpha_k) =\alpha_k \notag
\end{eqnarray}
with $\alpha_k$ the ratio of the wavenumbers. The ratios can be related to the geometric angles with the hat accent $\hat \alpha$ as illustrated in~\fref{fig:eddy-tilt}.  Note that the ratio of the wavenumbers $\alpha_k$ is opposite (or in geometric angle $\pi/2$ shifted) with respect to $\alpha_u$, because it reflects the angle of the corresponding wavevector (in a plane-wave-like approximation), rather than the geometric tilt of the ellipse-like structure. Although much of the theoretical considerations in~\cite{Fedorczak_2012,Fedorczak_2013} are based on $\alpha_k$, in the following $\alpha_u$ will be used instead, because it more naturally follows the intuitive geometric structure tilt mental image in real space.

With this formalism the local Reynolds stress can be expressed as
\begin{equation}
    \left\langle \tilde u_x\tilde u_y\right\rangle_t \approx -  \alpha_k \avg{k_y^2\tilde \phi^2}_t \approx \alpha_u \avg{\tilde u_x^2}_t
    \label{eq:RS-alphas}
\end{equation}
 where $\left\langle\dots\right\rangle_t$ represents the local (at a  given poloidal location) ensemble-only average\footnote{the subscript $t$ is motivated by the fact that it is typically estimated through a sufficiently long time average}.
 The original assumption of $k_x=k_y$ in~\cite{Eich_2021} is, therefore, equivalent to setting the average tilt to $\hat\alpha_u\sim -45^\circ$ as in the right image in~\fref{fig:eddy-tilt}. This would be consistent with a strong flow shear near the transition to  H-mode.

\begin{figure}
    \centering
    \includegraphics[width=\linewidth]{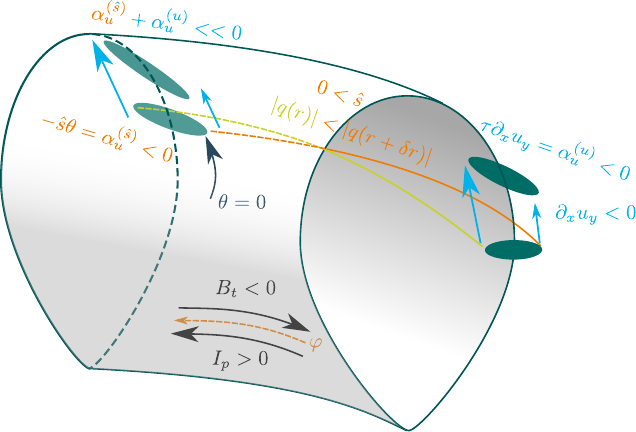}
    \caption{Schematic of tilting (represented by $\alpha_u$) of a structure originating at the outboard midplane due to  the flow shear $\partial_x u_y$ over a characteristic time $\tau$ (blue), and the magnetic shear $\hat s$ along field lines (orange). The displayed flux surface corresponds to a radial location just inside the separatrix, i.e. the outer shear layer. The displayed orientation of the magnetic field $B_t<0$ and the plasma current $I_p>0$ (negative helicity $q<0$) relative to the counter-clockwise toroidal angle $\varphi$ corresponds to the most common orientation in AUG.}
    \label{fig:tilting-schematic}
\end{figure}

However, here the theory of~\cite{Fedorczak_2013} comes in with a more elaborate relation for $\alpha_k$ also dependent on the magnetic shear. 
 In this theory the impact of both the flow shear as well as the magnetic shear along the field line on ballooning-like structures elongated along the field line (small $k_\parallel$) originating dominantly around the outer midplane (OMP) at poloidal angle $\theta=0$ is considered, as illustrated in~\fref{fig:tilting-schematic}. An initial radial wavenumber spectrum with zero mean  is assumed, so that on average (in ensemble sense) such structures will experience a tilt along the field line on the flux surface expressed through the formula~\cite{Fedorczak_2013} (recast here in terms of $\alpha_u$)
\begin{equation}
    \alpha_u(\theta) =  \underbrace{\tau\partial_x u_y}_{\alpha_{u}^{(u)}}   \underbrace{\vphantom{u_y}-\hat s \theta}_{\mathclap{\alpha_{u}^{(\hat s)}}} 
    \label{eq:kx-from-ky}
\end{equation}
with the typical eddy life time over which the structure experiences the flow shears  $\tau$, the mean shearing rate $\partial_x u_y$,  and the normalized magnetic shear parameter $\hat s=\partial \ln(q)/\partial \ln(x)=(x/q)\partial_x q$. For convenience the two tilt components are split into the flow-shear-induced tilt $\alpha_{u}^{(u)}$ and the  magnetic-shear-induced tilt $\alpha_{u}^{(\hat s)}$ (which incorporates the minus sign).

  The minus sign in $\alpha_{u}^{(\hat s)}$ reflects the fact that a positive magnetic shear over a counter-clockwise poloidal angle $\theta$ displacement tilts the structure clockwise.  Note that in Ref.~\cite{Fedorczak_2012} $\partial_x u_y$ is denoted as  $V_E'$ and the relation  \eqref{eq:kx-from-ky} is between $k_r$ and $k_\perp$. Indeed, a certain equivalence between the poloidal and bi-normal wavenumbers is assumed, which at the very edge of the plasma with a sufficiently high safety factor is likely reasonable.

The formula \eqref{eq:kx-from-ky} remains the same for any helicity (because $\hat s$ does not depend on the sign of $q$) and magnetic field orientation (incorporated in the flow shear $\partial_x u_y$ sign).
Broadly speaking, in a typical tokamak $\hat s\approx 2>0$ holds at the very edge of the confined plasma near the separatrix~\cite{Zohm.ch2}. Therefore, in a given magnetic equilibrium shape $\alpha_{u}^{(\hat s)}$ is expected to remain the same. On the other hand, the sign of the flow shearing rate and thus also $\alpha_{u}^{(u)}$ can be changed in the experiment by the toroidal field direction, and this can, therefore,  either act favorably or unfavorably relative to the magnetic shear. 

However, the favorable tilting superposition above the OMP shown in~\fref{fig:tilting-schematic} would become unfavorable below the OMP. Therefore, a certain poloidal asymmetry is required for a non-zero net effect.

To this end \cite{Fedorczak_2012} introduced another abstraction convenient for the coupling; The separation of the poloidal dependence of the potential fluctuation energy into a function $F_\phi$ representing only its poloidal dependence through  $\tilde \phi^2(\theta) = F^2_\phi(\theta) \tilde \phi^2(\theta=0)$. This separation facilitates the cancellation of $\tilde \phi^2$ in the approximations leading to \eref{eq:H-suppr-gamma-eff}. In~\cite{Fedorczak_2012} $k_y$ was assumed to be poloidally constant in a weak magnetic flux expansion limit, thus only the poloidal dependence of the potential fluctuation energy was considered. For greater generality here the actual radial velocity fluctuation energy $\tilde u_x^2\approx k_y^2\tilde \phi^2$ will be considered and the assumption of a poloidally constant $k_y$ will be further investigated in the following sections. Therefore, the $F$ function will be defined here as

\begin{equation}
    F^2(\theta):=\frac{\left\langle\tilde u_x^2\right\rangle_t(\theta)}{\langle \tilde u_x^2\rangle}; \qquad \left\langle F^2 \right\rangle = 1
    \label{eq:F-envelope}
\end{equation}

Combining this abstraction with \eref{eq:RS-alphas} results in

\begin{align}
\label{eq:RS-DALF-Fedorczak}
     \left\langle \tilde u_x\tilde u_y \right\rangle &\approx  \left\langle F^2\alpha_u\right\rangle   \langle \tilde u_x^2\rangle =: \alpha_\mathrm{RS}\langle \tilde u_x^2\rangle    \\
     \alpha_\mathrm{RS}\approx\frac{\left\langle \tilde u_x\tilde u_y \right\rangle}{\langle \tilde u_x^2\rangle} &\approx  \left\langle F^2\alpha_u\right\rangle = \underbrace{\left\langle F^2\alpha_u^{(u)}\right\rangle}_{\alpha_\mathrm{RS}^{(u)}} + \underbrace{\left\langle F^2\alpha_u^{(\hat s)}\right\rangle}_{\alpha_\mathrm{RS}^{(\hat s)}} \notag
\end{align}

where the proportionality constant $\alpha_\mathrm{RS}$ between the Reynolds stress and the average radial velocity fluctuations energy $\langle \tilde u_x^2\rangle $ is the average (now both in ensemble and flux surface sense) structure tilt weighted by the poloidal distribution of velocity energy $F^2$. The average tilt is again separated into components due to the magnetic shear $\alpha_\mathrm{RS}^{(\hat s)}$ and the flow shear $\alpha_\mathrm{RS}^{(u)}$. The form of~\eref{eq:RS-DALF-Fedorczak} with $\langle \tilde u_x^2\rangle$ factored out  is particularly convenient for the normalization of \eqref{eq:H-suppr-conceptual} by the total kinetic energy $E_{tk}\approx k_\perp^2\tilde\phi^2/2\approx k_y^2\tilde\phi^2$ with $k_y\approx k_x$ at the OMP as discussed later in~\sref{sec:sepos-RS-tilt}, because then the not well known $\tilde \phi^2$ cancels, highlighting how these theories are well suited for coupling.

However, it is important to note that the approximation of $E_{tk}$ through $\tilde\phi^2$  in \eref{eq:H-suppr-gamma-eff} also assumes $k_y=k_y(\theta=0)$ is constant on the flux surface, therefore, it will likely underestimate or overestimate $\langle \tilde u_x^2\rangle$ if $k_y=\mathrm{const}$ is not the case. Thus, for the purposes of the inclusion of $\alpha_\mathrm{RS}$ in  \eref{eq:H-suppr-gamma-eff} it is necessary to rescale it to that approximation as $\alpha_\mathrm{RS,0}$ in order to give the appropriate Reynolds stress estimate by
\begin{equation}
    \alpha_\mathrm{RS,0} = \frac{\avg{\tilde u_x^2}}{\avg{\tilde\phi^2}k_y^2(\theta=0)}\alpha_\mathrm{RS}
    \label{eq:alpha-RS-SepOS}
\end{equation}

As will be shown in more detail in \sref{sec:sepos-RS-tilt}, ultimately  $\alpha_\mathrm{RS,0}$ scales the left-hand-side term of \eqref{eq:H-suppr-gamma-eff}, according to the balance of $\alpha_\mathrm{RS,0}^{(\hat s)}$ and $\alpha_\mathrm{RS,0}^{(u)}$ determined by the combination of the X-point-induced asymmetry and flow shear orientation, respectively.

To determine what values of $\alpha_\mathrm{RS,0}$ are realistic and whether this model works, higher fidelity simulations will be investigated in the following \sref{sec:equilibrium-RS-model}.

One might be inclined to also deduce that a seemingly second point of coupling would be the mean poloidal flow $\langle u_y\rangle$ in the suppression criterion \eqref{eq:H-suppr-conceptual}, because its modification by the Reynolds stress is described by the poloidal momentum balance equation developed in~\cite{Fedorczak_2012} and slightly modified in~\cite{Peret2022}.

However, here the two theories employ somewhat different assumptions due to being applied in different circumstances. In~\cite{Eich_2021} the turbulence is assumed to be suppressed (in average sense) in H-mode conditions described by \eqref{eq:H-suppr-conceptual}, and thus the mean flow is approximated by the diamagnetic ion flow $\langle u_y\rangle\approx (\nabla p_i)/enB$ owing to the observation that in H-mode in both the favorable and unfavorable configurations, and in the absence of significant momentum input, the $E_r$ minimum (well depth) is typically observed to be close to this value set by neoclassical physics~\cite{Viezzer_2014,Plank_2023_PPCF}.

Therefore, in this model the Reynolds stress acts only as a ``conduit'' which enables the transfer of any free turbulent energy (represented on the right-hand side of \eqref{eq:H-suppr-conceptual}) generated due to steep pressure gradients back to the mean flow (set by the pressure gradient), thereby maintaining suppression.

\section{Equilibrium Reynolds stress asymmetry model}
\label{sec:equilibrium-RS-model}

To validate a model for $\alpha_\mathrm{RS}$ one would ideally use a full poloidal profile of the Reynolds stress measured in an experiment. While several measurements of the Reynolds stress in various devices  and using different diagnostic techniques exist, most have been done only in the vicinity of the OMP or did not cover a sufficiently dense poloidal profile to enable a flux-surface average estimate.

Therefore, one can attempt to validate against  the next best thing, which is a state-of-the-art turbulence simulation well matched with experimental observations. For this particular study the best option at the time of writing appeared to be the GRILLIX full-f fluid simulation of edge plasma in realistic X-point equilibrium geometry, validated and well matched against an AUG L-mode discharge thanks to the inclusion of a simple neutrals diffusion model~\cite{Zholobenko_2021-neutrals}. Similarly, the GRILLIX simulations of both favorable and unfavorable TCV L-mode discharges in the TCV-X21 dataset~\cite{TCV-X21} will be analyzed to investigate the difference between favorable and unfavorable configurations. It is however fair to note that in particular the AUG GRILLIX simulation has not fully matched the experimental radial electric field $E_r$ profile~\cite{Zholobenko_2021-Er}, therefore, the flow patterns discussed below could in principle be somewhat different in reality.

In the following quantities from the GRILLIX simulations will be shown averaged over the converged portion of a given simulation, i.e. they will be shown with the ensemble approximation average $\langle\dots\rangle_t$ applied. However, for simplicity this bracket operator will be omitted in the following, except when its application is not obvious.

\begin{figure}
    \centering
    \includegraphics[width=\linewidth]{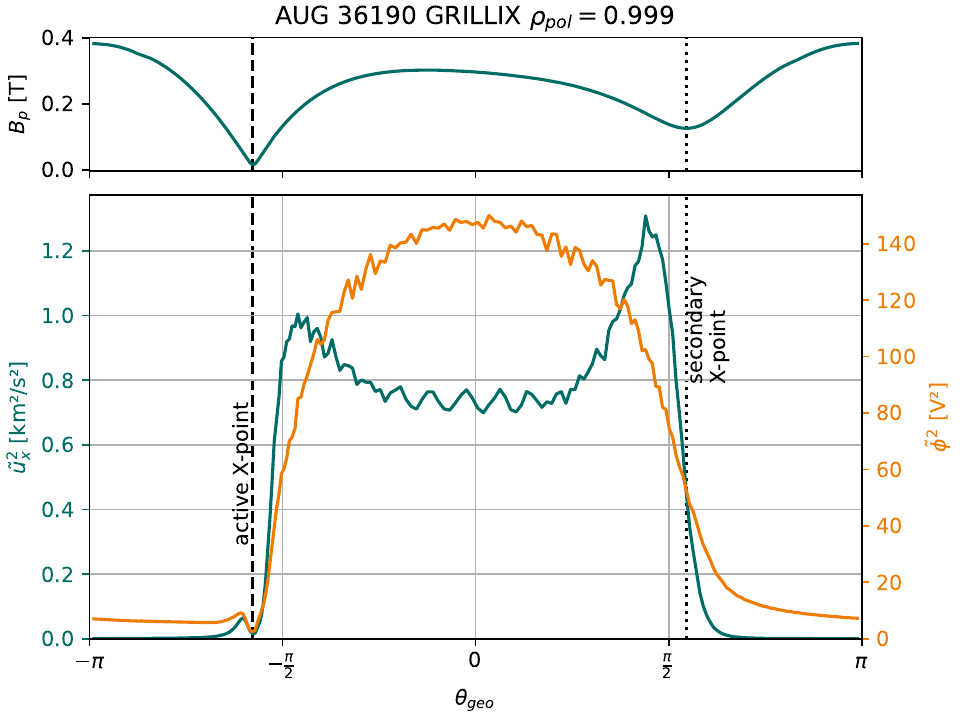}
    \caption{Radial velocity fluctuation energy $\avg{\tilde u_x^2}$ and potential fluctuation energy $\avg{\tilde \phi^2}$ profile in ASDEX Upgrade GRILLIX simulation of L-mode just a little inside the separatrix, shown with respect to the geometric angle $\theta_{geo}$. Locations in the proximity of the active or the secondary X-points are shown by dashed lines, corresponding to local minima of the poloidal magnetic field $B_p$.}
    \label{fig:F-geo}
\end{figure}

Since $F^2$ is assumed to be approximately an even function of $\theta$ (symmetric about the OMP) and $\alpha_u^{(\hat s)}$ approximately odd in $\theta$,  it is evident that for a non-zero Reynolds stress originating from the magnetic shear component the average $\langle F^2 \alpha_u^{(\hat s)}\rangle$ must be non-zero either due to a poloidal asymmetry of $F^2$ or $\alpha_u^{(\hat s)}$. In~\cite{Fedorczak_2013} $\alpha_u^{(\hat s)}$ was assumed to be a simple odd function of $\theta$ with a constant $\hat s$ on the flux surface, and thus the asymmetry had to come from  $F^2$. To this end $F^2$ was approximated by a poloidally periodic Gaussian centered around the OMP, representing a ballooned fluctuation drive envelope related to normal curvature, but truncated around the X-point  due to resistivity effects around the X-point. Initial investigation of the GRILLIX $F^2$ (both potential and velocity) for the reference simulation shown in~\fref{fig:F-geo} indeed suggests that this picture is reasonable. Nevertheless, it is also evident that the profiles of $\avg{\tilde u_x^2}$ and $\avg{\tilde \phi^2}$ do not have the same shape, thus the assumption of a constant $k_y$ is not valid.

When it comes to calculating a flux surface average of the Reynolds stress, it is, however, necessary to not look only at the geometric angle as was done in~\cite{Fedorczak_2012}. For this reason, poloidal profiles of quantities will be shown against the constant volume differential angle $\theta_V$ as defined for instance in~\cite{jardin2010computational},
because $\theta_V$ is the effective angle over which the true flux surface average is performed as
\begin{equation}
    \langle \dots \rangle = \frac{1}{\oint B_p^{-1} \mathrm{d}l_{pol}}\oint \dots \frac{\mathrm{d}l_{pol}}{B_p} = \frac{1}{2\pi}\oint \dots \mathrm{d}\theta_V
    \label{eq:theta-V}
\end{equation}
Therefore, poloidal profiles with respect to $\theta_V$ represent the distribution\footnote{The weighting by $1/B_p$ essentially represents flux expansion, because the flux surface average is actually a volume average  over an infinitesimally radially narrow (differential) volume around the flux surface.} of contributions to the flux surface average.  The reference location $\theta_V=0$ is associated with the OMP where interchange-like structures are assumed to dominantly originate. Essentially, $\theta_V$ is ``stretched out'' with respect to the geometric angle around locations of low poloidal field.

\begin{figure}[htb]
    \centering
    \includegraphics[width=\linewidth,height=0.8\textheight,keepaspectratio]{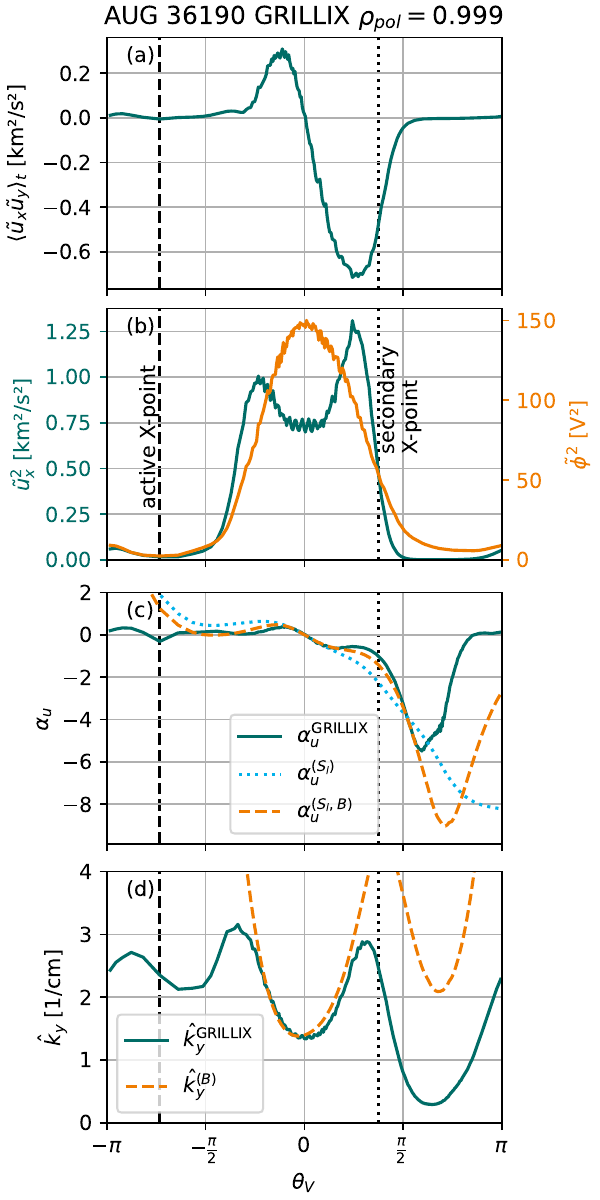}
    \caption{Poloidal profiles of the Reynolds stress $\langle \tilde u_x\tilde u_y\rangle_t$ (a), the radial fluctuation velocity $\tilde u_x^2$ and potential fluctuation $\tilde\phi^2$ energies (b), average effective tilt structure angle $\alpha_u$ (c) and effective poloidal wavenumber $\hat k_y$ at the reference radial location $\rho_{pol}=0.999$ in GRILLIX simulation of AUG discharge 36190~\cite{Zholobenko_2021-neutrals}. }
    \label{fig:F-model}
\end{figure}

In \fref{fig:F-model} the poloidal  profiles of the discussed quantities at $\rho_{pol}=0.999$ are shown. This radial location just a little inside the separatrix is chosen, because it corresponds to the model in \eqref{eq:H-suppr-conceptual}, and also conveniently there is little flow shear $\partial_x u_y \sim 0$ which enables to study the impact of the magnetic shear in a somewhat isolated manner. 

Firstly, it is evident in \fref{fig:F-model} panel (a) that the Reynolds stress profile displays a significant asymmetry, particularly in the peak-to-peak magnitudes. In order to investigate to what extent this asymmetry is due to asymmetries in $F^2$ the profiles of the potential fluctuation energy $\tilde\phi^2$ and the radial velocity fluctuation energy $\tilde u_x^2$, corresponding to the un-normalized $F^2$ profile shape in the case where no poloidal variation of $k_y/B$ is neglected or included, respectively, are shown in  \fref{fig:F-model} panel (b). Interestingly, in terms of the flux-surface angle $\theta_V$ the profiles appear to be quite symmetric compared with the Reynolds stress asymmetry. Particularly in terms of their poloidal extent, the truncation by the primary X-point below the OMP as shown in~\fref{fig:F-geo} and assumed in~\cite{Fedorczak_2012} is not as pronounced anymore.   The reason for this approximate symmetry in $\theta_V$ particularly near the X-point regions is that the lower $B_p$ in these regions compensates through \eqref{eq:theta-V} for the truncation asymmetry in the real-space geometry. The small difference between the peak heights of $\tilde u_x^2$ would account only partly for the much larger difference in the magnitudes of the Reynolds stress peaks.

Therefore, the pronounced Reynolds stress asymmetry appears to be not just due to $F^2$ in this case. Instead it seems the dominant origin of the asymmetry has a different reason. Inspired by the explanation offered in~\cite{StaeblerPoP2011} that it is the asymmetry of the local magnetic shear, an attempt to generalize the $\alpha_u^{(\hat s)}\approx -\hat s \theta$ expression is made in the following.

The generalization of the magnetic shear of field lines is
\begin{equation}
    -\hat s \theta\rightarrow -\int\limits_0^{l(\theta)} S_{l,g} \mathrm{d}l(\theta) =: \alpha_u^{(S_l)}
    \label{eq:alpha-k-S_l}
\end{equation}
where the geometric local magnetic shear $S_{l,g}$ is integrated along the field line length $l$ up to angle $\theta$ relative to $\theta=0$ at the OMP. It is directly related to the local magnetic shear  $S_l$ defined in~\cite{Carthy_2013} through $S_{l,g}=\mathrm{sign}(q)S_l$. Here the sign of the safety factor $q$ carries the signs of the magnetic field vectors relative to the fixed typical geometry orientation, i.e. counter-clockwise toroidal and poloidal angles, because  in~\cite{Carthy_2013} $S_l$ is defined with respect to the magnetic field vector orientation and will change signs depending on the field orientation, whereas for the following geometric considerations $S_{l,g}$ is relative to the aforementioned fixed geometry orientation. This sign cancellation by $q$ is also illustrated in the high-aspect-ratio symmetry limit $ S_l q R \rightarrow \hat s$.

In \fref{fig:F-model} panel (c) the effective angle tilt profile $\alpha_u^\mathrm{GRILLIX}=\langle \tilde u_x\tilde u_y\rangle_t / \tilde u_x^2$ in the simulation is compared with the tilt angle $\alpha_u^{(S_l)}$ calculated through \eqref{eq:alpha-k-S_l} with the local magnetic shear. The agreement is already quite reasonable in terms of roughly capturing the asymmetry, which is due to the local magnetic shear approaching 0 towards the active X-point coming from the OMP.  Nevertheless, the agreement can be further improved by considering the poloidal variation of $k_y$ in the following.

An effective poloidal wavenumber is shown in \fref{fig:F-model} panel (d), which is evaluated from the two profiles in panel (b) as $\hat k_y^\mathrm{GRILLIX}=\sqrt{\tilde u_x^2/(\tilde\phi/B_t)^2}$ based on the $E\times B$ velocity prescription. The local toroidal magnetic field strength $B_t$ is used since AUG is not a high-aspect-ratio tokamak and $B_t$ varies poloidally on a flux surface. The toroidal field is used in here instead of the total field one would expect in an $E\times B$ velocity, because GRILLIX approximates the perpendicular gradient in the poloidal plane only. 
 
It is evident that $\hat k_y^\mathrm{GRILLIX}$ significantly increases in the proximity of X-points relative to its value near the OMP. The decrease further beyond the X-points is not truly representative, because both $\tilde u_x^2$ and $\tilde\phi^2$ vanish towards 0 (the latter slower) and there is not much turbulence to speak of, hence their ratio is not entirely meaningful in those regions.  While approximating $\hat k_y$ analytically is generally difficult due to the non-linear nature of the studied turbulence regime,
the proximity of the increase in $\hat k_y$  to the X-points motivates the following argument based on the magnetic field line scales.

Let us assume that through field line helicity the poloidal wavenumber $k_y$ is related to a parallel wavenumber $k_\parallel$. Further assuming there is a dominant parallel wavenumber, perhaps related to a critical normalized gradient $\alpha_c$ as assumed in~\cite{Eich_2021}, we can relate the phase increment along a field line increment $\Delta l$ with the same increment projected onto a poloidal arclength increment $\Delta l_{pol}$ as $k_\parallel\Delta l \approx k_y \Delta l_{pol}$. Since the ratio of the length increments is given by the field line pitch angle, the poloidal wavenumber can be approximated as

\begin{equation}
    \hat k_y^{(B)} :=  \hat k_{\parallel}\frac{B}{B_p}
    \label{eq:k-y-B}
\end{equation}
where $\hat k_\parallel$ is a reference parallel wavenumber. Setting $\hat k_\parallel$ to agree with $\hat k_y^\mathrm{GRILLIX}$ at the OMP the comparision of $\hat k_y^{(B)}$ with $\hat k_y^\mathrm{GRILLIX}$ in \fref{fig:F-model} panel (d) shows that this approximation indeed captures the trend of increasing $k_y$ towards the X-points rather well. Naturally, there is a big disagreement once the estimation of  $\hat k_y^\mathrm{GRILLIX}$ is meaningless due to the vanishing turbulence levels.

Considering that \eref{eq:kx-from-ky} is derived in eikonal formulation in~\cite{Fedorczak_2012} effectively through an increment of the radial wavenumber $\Delta k_x = k_y \hat s   \Delta \theta$, \eref{eq:alpha-k-S_l} can be generalized using the $k_y$ approximation from \eref{eq:k-y-B} as 
\begin{equation}
    -\hat s \theta = -\frac{1}{k_y} \int k_y \hat s  \mathrm{d}\theta\rightarrow -\frac{1}{\frac{B}{B_p}}\int\limits_0^{l(\theta)}\frac{B}{B_p} S_{l,g} \mathrm{d}l(\theta) =: \alpha_u^{(S_l,B)}
    \label{eq:alpha-k-S_l-B}
\end{equation}
As can be seen in \fref{fig:F-model} panel (c) this refined approximation $\alpha_u^{(S_l,B)}$ of the magnetic-shear-induced tilt agrees remarkably well with the simulated effective tilt $\alpha_u^\mathrm{GRILLIX}$. This result points towards the magnetic shear asymmetry as being the dominant origin of the Reynolds stress asymmetry, since it accounts for the large asymmetry in $\alpha_u^\mathrm{GRILLIX}$, which in turn accounts for the main asymmetry in the Reynolds stress. 

Altogether, the average tilt calculated from the profiles shown in~\fref{fig:F-model} following \eref{eq:RS-DALF-Fedorczak} (i.e. the ratio of the Reynolds stress and radial fluctuation energy flux surface integrals) gives $\alpha_\mathrm{RS}\approx -0.24$. When the variation of the poloidal wavenumber is not taken into account and the result is rescaled according to \eref{eq:alpha-RS-SepOS}, the average tilt is $\alpha_\mathrm{RS,0}\approx -0.35$.

\begin{figure}
    \centering
    \includegraphics[width=\linewidth,height=0.8\textheight,keepaspectratio]{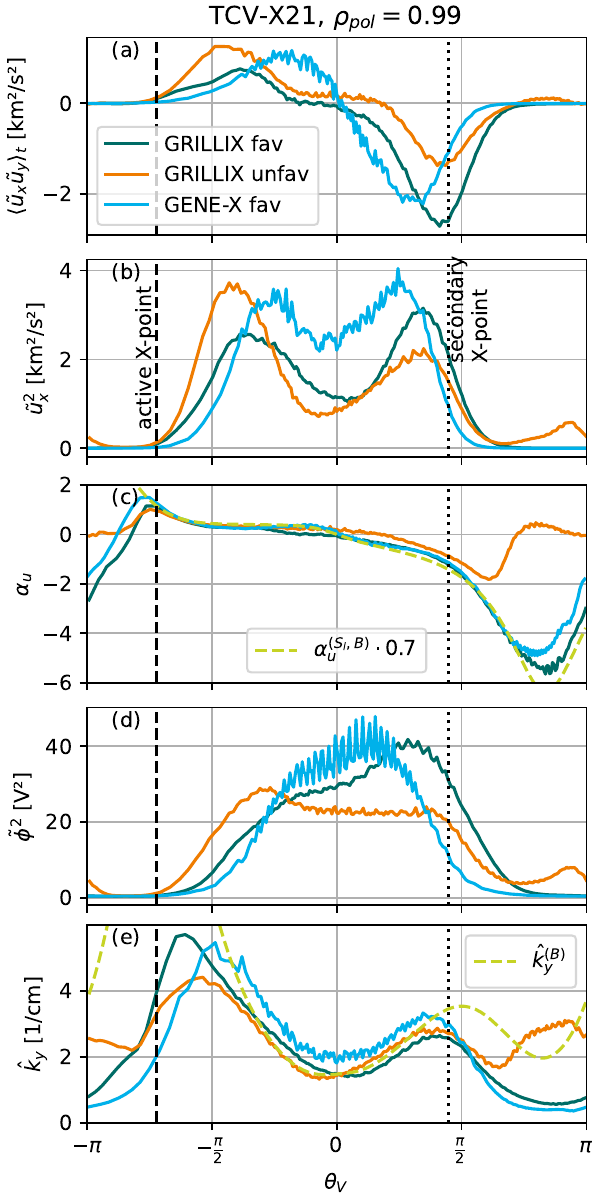}
    \caption{Poloidal profiles of the Reynolds stress $\langle \tilde u_x\tilde u_y\rangle$ (a), the radial fluctuation velocity $\tilde u_x^2$  energies (b), average effective tilt structure angle $\alpha_u$ (c), potential fluctuation energy $\tilde\phi^2$ (d), and effective poloidal wavenumber $\hat k_y$ (e)  at the reference radial location $\rho_{pol}=0.99$ in GRILLIX simulations of favorable and unfavorable TCV-X21 cases~\cite{TCV-X21} as well as the GENE-X favorable case~\cite{TCV-X21-GENE-X}. }
    \label{fig:TCV-X21-GRILLIX}
\end{figure}

The next step is to investigate whether this magnetic-shear-induced asymmetry remains the same between the favorable and unfavorable configuration when only the magnetic field direction is changed. To this end a similar analysis has been applied to the TCV-X21 favorable and unfavorable cases and the results are displayed in~\fref{fig:TCV-X21-GRILLIX}. Additionally, the analysis was performed also for the corresponding simulation of the favorable case with GENE-X~\cite{TCV-X21-GENE-X}. For these cases the radial reference location a little deeper inside $\rho_{pol}=0.99$ is shown in order to to highlight the impact of the finite flow shear.

It is clearly seen in \fref{fig:TCV-X21-GRILLIX} panel (a) that the Reynolds stress profiles  significantly differ between the favorable and unfavorable GRILLIX cases. However, their shape is rather similar, the main difference appears to be mainly of an offset-like nature. The favorable GENE-X case is rather different, though still has a sine-like dependence.

Firstly, the asymmetry of the of the potential and radial velocity energy fluctuations are investigated to determine if they are the dominant source of the Reynolds stress asymmetry. Clearly, the asymmetry in the potential and radial velocity fluctuation envelopes in panels (b) and (d), respectively, between the two GRILLIX cases is quite significant. But this is not due to an X-point truncation (the X-point location does not change) as assumed in~\cite{Fedorczak_2012}, but rather a ``skewing'' of the whole envelope. This might suggest that additional effects also play a role in the Reynolds stress asymmetry. In~\cite{Fedorczak_2012} it was proposed that the envelope could poloidally move  due to the Bloch shift of the interchange growth rate to the poloidal location where the wavenumbers are the smallest, i.e. to the location with $\alpha_u=0$. However, the peaks of $\tilde \phi^2$ in \fref{fig:TCV-X21-GRILLIX} panel (d) go in the opposite direction, rather following the flow direction.

To further investigate this additional $\tilde \phi^2$ asymmetry, a similar analysis was also done for the corresponding favorable GENE-X simulation~\cite{TCV-X21-GENE-X}. There the asymmetry of $\tilde \phi^2$ was found to be significantly smaller. Furthermore, the envelope is even more ballooned, likely due to the presence of trapped electron mode turbulence (TEM). Inclusion of a Landau fluid closure into GRILLIX~\cite{PitzalLFinPreparation} to get a more realistic heat flux model, i.e. taking it closer to GENE-X, also leads to more symmetric profiles. This suggests that the large envelope asymmetry may not be entirely realistic and could be down to the limitations of the Braginskii model used in those GRILLIX simulations.

Therefore, since the $\tilde \phi^2$ asymmetry is likely  unrealistic, we turn our attention again to the estimated average tilt $\alpha_u$ in \fref{fig:TCV-X21-GRILLIX} panel (c). The general asymmetry shape of $\alpha_u^\mathrm{GRILLIX,GENE-X}$ besides the offset  appears to be the same in both the favorable and unfavorable cases. The $\alpha_u^{(S_l,B)}$ formula again captures the $\alpha_u^\mathrm{GRILLIX,GENE-X}$ asymmetry shape quite well, up to a numerical factor of 0.7. This overestimation (in absolute value) by $\alpha_u^{(S_l,B)}$  can be attributed to the effective correlation $\rho_{corr}=\langle \tilde u_x\tilde u_y\rangle_t/\sqrt{\tilde u_x^2 \tilde u_y^2}$ being significantly below 1 especially around the OMP, i.e. to the departure from the near-perfect linear correlation of velocities assumed in \eref{eq:alpha-k}. This may be due to to the turbulence regime being different compared to the AUG case and/or could be a consequence of the smaller size of the TCV geometry, where the ratio of the turbulence structures size and the equilibrium geometry scales may be larger than in AUG. 

Nevertheless, the main  asymmetry in the shape of the Reynolds stress profile in all cases shape again appears to be mainly due to the magnetic shear asymmetry, as was the case with the AUG simulation.

The offset-like difference seen most clearly above the OMP between $\alpha_u^\mathrm{GRILLIX}$ in the favorable and unfavorable GRILLIX cases  is related to the finite flow shear and consistent with \eref{eq:alpha-k}, where $\alpha_u^{(u)}$ is the offset which changes sign according to the shear flow orientation.   The offset between the two cases is also very small near the active X-point, which can be understood by the local shearing rate vanishing $\partial_x u_y \rightarrow 0$ due to strong flux expansion, but the lifetime $\tau$ not approaching infinity at the same time since even in this region turbulent structures have a finite average lifetime, i.e. $\alpha_u^{(u)}\rightarrow 0$ in this location.

Looking at the average tilts, one finds $\alpha_\mathrm{RS}\approx -0.27$ and $\alpha_\mathrm{RS,0}\approx -0.85$ for the favorable case, and $\alpha_\mathrm{RS}\approx 0.07$ and $\alpha_\mathrm{RS,0}\approx 0.26$ for the unfavorable case, respectively. Assuming that the magnetic $\alpha_\mathrm{RS,0}^{(\hat s)}$ and flow $\alpha_\mathrm{RS,0}^{(u)}$ are in magnitude (but $\alpha_\mathrm{RS,0}^{(u)}$ changes sign) the same in both cases, following \eref{eq:RS-DALF-Fedorczak} one can estimate $\alpha_\mathrm{RS,0}^{(\hat s)}\approx -0.3$ and $|\alpha_\mathrm{RS,0}^{(u)}|\approx 0.5$. 

 For completeness let us state that for the favorable GENE-X simulation $\alpha_\mathrm{RS}\approx -0.11$ and $\alpha_\mathrm{RS,0}\approx -0.31$ is found. Since the average tilt profile $\alpha_u^\mathrm{GENE-X}$ in \fref{fig:TCV-X21-GRILLIX} panel (c) crosses 0 close to $\theta_V=0$, it appears it is less affected by the flow shear component. Therefore, it can be viewed a being more representative of mainly the magnetic shear component, similarly as appeared to be the case in the AUG simulation discussed before. 

Therefore, altogether both the AUG GRILLIX and GENE-X TCV cases, as well as the estimate from the difference of the two favorable/unfavorable GRILLIX TCV cases,   appear to point to the average tilt  due to the magnetic shear at the very edge of the plasma being about $\alpha_\mathrm{RS,0}^{(\hat s)}\approx -0.3$ in a LSN configuration.

\section{Combined separatrix H-mode sustainment model}
\label{sec:sepos-RS-tilt}

For the final combination of \eqref{eq:H-suppr-conceptual} and \eqref{eq:RS-DALF-Fedorczak} it is necessary to approximate $\tau\partial_x u_y$. Since \eqref{eq:H-suppr-conceptual} aims to describe H-mode suppression in the vicinity of the OMP, it is reasonable to assume $\tau\partial_x u_y\sim 1$ as in typical turbulence shear-suppression criteria $1/\tau\sim \partial_x u_y$. However, this is likely to be fulfilled only around the OMP, because further away towards the X-points the $E\times B$ flow (assuming $\langle\phi\rangle_t$ is close to a flux function) and the corresponding velocity shear is weaker due to flux expansion, as was already suggested by the analysis of \fref{fig:TCV-X21-GRILLIX}. Therefore, likely $|\alpha_\mathrm{RS}^{(u)}|=|\avg{ F^2 \tau\partial_x u_y}| \lessapprox 1$. 

On the other hand, because around the OMP likely $|\tau\partial_x u_y|\sim 1$  while $\alpha_u^{(\hat s)}\approx 0$, $|\alpha_u(\theta=0)|\approx 1$ is a reasonable assumption in the state of H-mode suppression and so is the original approximation of $k_x^2=\alpha_u^2 k_y^2\approx k_y^2:=k_\mathrm{ES}^2$ for the other terms in \eref{eq:H-suppr-conceptual}.

Specifically, the $\partial_x$ gradient of the Reynolds stress in \eqref{eq:H-suppr-conceptual} giving the Reynolds force was in~\cite{Eich_2021}\footnote{In~\cite{Eich_2021} the actual approximation used the unnecessary Fourier analysis ansatz with the imaginary unit $\partial_x \approx i k_\mathrm{ES}$ which then necessitated another imaginary unit in the mean velocity $\avg{u_y}=-i \tau_i \Lambda_{p_i}$. Through the full treatment of the velocity orientation in \eref{eq:RS-suppr-partial-int} and \eref{eq:u_y-general} this argument is no longer required.}  approximated by the typical turbulence scale $\partial_x \approx k_\mathrm{ES}$. Since this operator acts already at the OMP on the average Reynolds stress, this approximation is retained. Similarly, the total turbulent kinetic energy at the OMP remains $E_{tk}=\tilde u^2/2\approx(k_x^2+k_y^2) \tilde\phi^2/2 =k_y^2\tilde \phi^2$, which is then equivalent to the radial velocity energy. In principle this radial velocity energy $\avg{\tilde u_x^2}_t(\theta=0)$ does not have to be equal to the flux-surface average $\avg{\tilde u_x^2}$, but the analysis in \sref{sec:equilibrium-RS-model} suggests that they are sufficiently close, therefore, the original approximation is also retained. 

The mean flow is again approximated as the ion diamagnetic velocity in the electron diamagnetic (hence no negative sign) direction. To capture its poloidal orientation along $\theta$, as shown in \fref{fig:tilting-schematic}, dependent on the magnetic field orientation, the approximation is generalized as
\begin{equation}
    \avg{u_y}\approx \left(c_s \frac{\rho_s}{\lambda_{p_e}}\right) \left(\frac{\nabla_x p_i \times \vec B }{eB^2}\right)=-\mathrm{sign}(B_t)\tau_i\Lambda_{p_i} 
    \label{eq:u_y-general}
\end{equation}
where the model normalization (for details see~\cite{Eich_2021}) is done by the cold-ion sound speed $c_s=\sqrt{T_e/m_i}$ and the drift scale $\rho_s=\sqrt{T_e m_i}/eB$,  and the positive sign of the toroidal magnetic field, $\mathrm{sign}(B_t)>0$, corresponds to the counter-clockwise orientation as viewed from above, i.e. along the toroidal angle $\varphi$ shown in \fref{fig:tilting-schematic}.

This approximations implies that the toroidal rotation component is assumed to have no significant radial shear, because $\avg{u_y}$ approximates the shearing rate as will be explained in the following. The approximation could also possibly somewhat overestimate the mean flow in L-mode conditions. Nevertheless, even with this caveat the criterion \eref{eq:H-suppr-conceptual} is still able to separate L-mode and H-mode conditions sufficiently well as seen in~\fref{fig:H-mode-suppr-fav-unfav}.

In order to get the signs of the suppression term dependent on the field and flow orientations right, it is necessary to revisit the suppression energy transfer term on the left-hand-side of \eref{eq:H-suppr-conceptual}. The  term $\langle u_y\rangle \partial_x \langle \tilde u_x\tilde u_y\rangle $ describing the net transfer of energy to the flow is often called the ``Reynolds work''. Although ``Reynolds work'' is the customary name, it is actually a transfer of energy, so it corresponds in terms of units to a power (per mass). This term is  an approximation of the actual suppression transfer of only turbulence energy to the flow $ \langle \tilde u_x\tilde u_y\rangle \partial_x \langle u_y\rangle$ as derived in~\cite{Manz2012PoP-suppr-model}.  The approximation holds under the assumption that the local sources of flow momentum are negligible $\partial_x\left( \langle \tilde u_x\tilde u_y\rangle \langle u_y\rangle\right)\approx 0$ as shown in the following partial integration 
\begin{eqnarray}
    \langle \tilde u_x\tilde u_y\rangle \partial_x \langle u_y\rangle &=\partial_x\left( \langle \tilde u_x\tilde u_y\rangle \langle u_y\rangle\right) - \langle u_y\rangle \partial_x \langle \tilde u_x\tilde  u_y\rangle \nonumber \\
    &\approx - \langle u_y\rangle \partial_x \langle \tilde u_x\tilde  u_y\rangle  
    \label{eq:RS-suppr-partial-int}
\end{eqnarray}
Most importantly, the last term has a minus sign  when the flow orientation is fully taken into account. For the approximation of the radial derivative of the Reynolds stress again the typical turbulence scale is used, because the Reynolds stress in the outer shear layer is the largest (in absolute value) near the separatrix with the largest shear and lowest (in absolute value) deeper inside towards the $E_r$ well with vanishing shear.

Altogether, the normalized stabilizing term\footnote{except for  the non-adiabaticity correction $1/(1+\delta_{\phi,p_e}^2)$. Since it is significant only at very high densities,  it will not be written explicitly in the following in the interest of clarity} on the left-hand side of \eref{eq:H-suppr-gamma-eff} is finally approximated as
\begin{equation}
    \gamma_\mathrm{suppr,eff} = \frac{- \langle u_y\rangle \partial_x \langle \tilde u_x\tilde  u_y\rangle}{E_{tk}}  =\tau_i\Lambda_{p_i} k_\mathrm{ES} \underbrace{ \mathrm{sign}(B_t) \alpha_\mathrm{RS,0}}_{\alpha_{\mathrm{RS},0,B}}
    \label{eq:gamma-suppr-new}
\end{equation}
which is nearly the same as in \eref{eq:H-suppr-gamma-eff} except for the factor $\alpha_{\mathrm{RS},0,B}:=\mathrm{sign}(B_t) \alpha_\mathrm{RS,0}$. This factor now captures the difference between the favorable and unfavorable $\nabla B$ drift directions. This is illustrated in \tref{tab:alpha-RS-sepos} for the typical experimental cases in AUG. For its construction it is important to recall that $\alpha_\mathrm{RS,0}^{(u)}$ depends on the orientation of the magnetic field as $\alpha_\mathrm{RS,0}^{(u)}\propto \partial_x \avg{u_y}\propto \mathrm{sign}(B_t)$, because at the outer shear layer the flow is expected to be (in absolute value) greater deeper inside, and approaching 0 near the separatrix, while $\alpha_\mathrm{RS,0}^{(\hat s)}$ is determined by the X-point location in the equilibrium. 

While $\alpha_\mathrm{RS,0}^{(u)}$ is not actually known, it can be inferred in the following from the fact that $\alpha_{\mathrm{RS},0,B} \approx 1$ holds robustly for the favorable cases: Assuming $\alpha_\mathrm{RS,0}^{(\hat s)}\approx -0.3$ estimated from GRILLIX and GENE-X in \sref{sec:equilibrium-RS-model} for the LSN configuration is sufficiently general, this means the shear flow component is approximately $\alpha_\mathrm{RS,0}^{(u)}\approx -0.7$ in this case where in LSN the favorable magnetic field orientation is $B_t<0$ This is consistent with the expectation that it will be, in absolute value, somewhat lower than 1. For the unfavorable LSN case it is assumed that $\alpha_\mathrm{RS,0}^{(\hat s)}$ remains the same, but the shear flow now changes sign as does, therefore, $\alpha_\mathrm{RS,0}^{(u)}\approx + 0.7$. Its magnitude is assumed to be the same at the point of H-mode turbulence suppression, which is consistent with experimental observations~\cite{Plank_Er_2023}. This gives a factor of $\alpha_{\mathrm{RS},0,B}\approx 0.4$, which is in very good agreement with the experimental observations in~\sref{sec:sepos-empirical}. In other words,  if the suppression term is by a factor $0.4=1/2.5$ weaker, the original stabilizing term in \eref{eq:H-suppr-gamma-eff} would indeed need to be 2.5 times larger. 

For the USN cases the shear-induced tilt component is expected to change sign by symmetry but retains its value since most USN AUG equilibria are quite comparable to up-down mirrored LSN equilibria, i.e.  $\alpha_\mathrm{RS,0}^{(\hat s)}\approx +0.3$. The resulting value of the $\alpha_{\mathrm{RS},0,B}$ factor then shows which $B_t$ orientation is the favorable or unfavorable one in this configuration as shown in \tref{tab:alpha-RS-sepos}.

\begin{table*}[htb]
    \centering
    \begin{tabular}{rrr|rrr}
         X-point & $\mathrm{sign}(B_t)$  & $\nabla B$ drift  & $\alpha_\mathrm{RS,0}^{(\hat s)}$ & $\alpha_\mathrm{RS,0}^{(u)}$ & $\mathrm{sign}(B_t)\alpha_\mathrm{RS,0}$ \\\hline
         LSN & -1  & favorable  & -0.3 & -0.7 & 1\\
         LSN & +1 & unfavorable  & -0.3 & +0.7 & 0.4\\
         USN & -1  & unfavorable  & +0.3 & -0.7 & 0.4\\
         USN & +1 & favorable  & +0.3 & +0.7 & 1\\
    \end{tabular}
    \caption{Comparison of the favorable vs. unfavorable configuration from the perspective of the global fields orientation with the average tilt factor.}
    \label{tab:alpha-RS-sepos}
\end{table*}

Altogether, the extended model with $\alpha_{\mathrm{RS},0,B}$ correctly describes the favorable and unfavorable behavior consistently with the usual $\nabla B$ drift orientation, though it does not have a direct physical connection with the $\nabla B$ drift. It also does not depend on the sign of the helicity, consistently with the $\nabla B$ drift behavior. 

\section{Consistency with  $E_r$ profile measurements}
\label{sec:Er-L-mode-consistency}

In order to attempt to recover the difference in radial electric fields in L-mode in the different configurations reported in~\cite{Plank_Er_2023}, one possibility is to attempt to use the momentum balance equation from~\cite{Fedorczak_2012} or even its combination with the spectral filament model in~\cite{Peret2022}. However, those models have several free parameters for which first-principles-based estimates are difficult to obtain, such as an ad-hoc poloidal momentum diffusivity or the width of the $\mathrm{tanh}(x)$-like gate function simulating the transition to SOL conditions. Furthermore,  the $\alpha_\mathrm{RS}^{(\hat s)}$ estimates obtained from the simulations are almost three times lower than those used in~\cite{Peret2022}. These various parameters  can have a significant impact on the final profile. Other parameters such as the average radial velocity fluctuation energy scale $\avg{u_x^2}$, which are not directly measured, can be estimated, for instance using the sheared spectral filament model as in~\cite{Peret2022}, but that model is based  on interchange physics only and likely is not entirely appropriate for lower collisionalities with significant drift-wave presence.

Therefore, an attempt to recover the observed $E_r$ difference is made directly based on the SepOS framework with known quantities. The effective  bi-normal flow velocity related to  $E_r$ can be obtained from the suppression term on the left-hand side of \eref{eq:H-suppr-conceptual} if it is normalized by the ``Reynolds force'' (per mass) $\partial_x \langle \tilde u_x\tilde  u_y\rangle$ (neglecting the non-adiabaticity correction for clarity again). In terms of dimensions analysis the suppression energy transfer rate - i.e. power (per mass) divided by a force (per mass) indeed gives a velocity $v=P/F$.   Expressed through the updated suppression criterion \eref{eq:gamma-suppr-new} this gives
\begin{equation}
     - \langle u_y\rangle =  \frac{\gamma_\mathrm{suppr,eff}E_{tk}}{\partial_x \langle \tilde u_x\tilde  u_y\rangle}  =\frac{\gamma_\mathrm{suppr,eff}}{ k_\mathrm{ES} \alpha_\mathrm{RS,0,B}}
    \label{eq:u_y-from-suppr}
\end{equation}

In order to understand to which location this mean flow $\langle u_y\rangle$ corresponds, it is 
illustrative to rewrite \eref{eq:u_y-from-suppr} as
\begin{equation}
    \frac{0-  \langle u_y\rangle }{1/(\alpha_\mathrm{RS,0,B}k_\mathrm{ES})} \approx \partial_x \langle u_y\rangle \propto  \gamma_\mathrm{suppr,eff}
\end{equation}
which expresses the implicit assumptions involved: Firstly, $\langle u_y\rangle$ actually approximates the true shearing rate $\partial_x \langle u_y\rangle$ in \eref{eq:RS-suppr-partial-int} which implicitly assumes that $\langle u_y\rangle\approx 0$ near the boundary.  Secondly, the radial gradient length is assumed to again be comparable to the turbulence scale $\partial_r \propto k_\mathrm{ES}$ as before. But the prefactor $\alpha_\mathrm{RS,0,B}$ effectively makes the gradient scale similar near the L-H boundary in both the favorable and unfavorable configurations, because in the unfavorable case at the higher temperature near the L-H boundary $k_\mathrm{ES}\propto \sqrt{\beta_e}$ is larger, but the smaller $\alpha_\mathrm{RS,0,B}$ compensates that. This roughly corresponds to the $E_r$ well width being roughly the same in both considerations, in agreement with experimental observations~\cite{Plank_Er_2023}. Finally, the suppression rate $\gamma_\mathrm{suppr,eff}$ is expected to approximate (at least in terms of its scaling)   the ``average'' shearing rate.
From this rough formula it is evident that $\langle u_y\rangle$ actually corresponds to a radial location deeper inside. Since $\gamma_\mathrm{suppr,eff}$ is assumed to be an  approximation of the ``average'' shearing rate across the outer shear layer, it follows  that $\langle u_y\rangle$  should correspond roughly to the $E_r$ minimum.

Further assuming that the background toroidal flow component is either small or is not significantly sheared (consistent with observations in~\cite{Plank_Er_2023}) and can thus be subtracted as an offset,  the electric field (without the toroidal component) can be estimated as $E_r\approx u_y B_t$, and removing the normalization, the (negative) $E_r$ minimum estimate at the OMP from the SepOS model can be expressed as
 \begin{equation}
     - \mathrm{min}(E_r^\mathrm{OMP}) \approx \frac{T_{e,sep}}{e\lambda_{p_e}/f_x} \frac{R_\mathrm{geo}}{R_\mathrm{geo} + a_\mathrm{geo}} \frac{\gamma_\mathrm{suppr,eff}}{ k_\mathrm{ES} \alpha_\mathrm{RS,0,B}}
    \label{eq:E_r-from-suppr}
\end{equation}
where, additionally, the flux expansion $f_x\sim 1.8$  accounts for the narrow gradient length at the OMP and the $R_\mathrm{geo}/(R_\mathrm{geo} + a_\mathrm{geo})$ rescales to the toroidal field $B_t$ to the OMP relative to the reference $B_t$ used in the normalization.
It may seem strange to use the separatrix temperature $T_{e,sep}$ for the estimation of the $E_r$ minimim deeper inside. One might expect some extrapolation deeper inside. But it is there  in fact  simply because $T_{e,sep}$ is  used for the normalization. In cases with significant, but insignificantly sheared, toroidal  rotation  \eref{eq:E_r-from-suppr} represents rather $ - \mathrm{min}(E_r^\mathrm{OMP}+v_i\times B)$.

The $E_r$ well minimum is in experiments  typically taken as a proxy for the shearing rate in cases where toroidal rotation is insignificant~\cite{Viezzer_2014}. In other words, here the SepOS approximation goes the other way than in the experiment.

To simulate a power scan from L-mode towards H-mode at fixed density and global parameters, the separatrix temperature $T_{e,sep}$ is virtually scanned towards the expected temperature at the L-H boundary $T_\mathrm{e,LH}(n_e)$ as a solution of \eref{eq:H-suppr-gamma-eff} (i.e. the blue line in \fref{fig:H-mode-suppr-fav-unfav}). To estimate the $E_r$ minimum in L-mode following \eref{eq:E_r-from-suppr} it is necessary to first approximate $\gamma_\mathrm{suppr,eff}$ in such conditions.
This can be done by approximating the ratio of the suppression and production rates $\gamma_\mathrm{suppr,eff}/ \gamma_\mathrm{prod,eff}$ by the proximity to the L-H boundary as
\begin{equation}
    \frac{\gamma_\mathrm{suppr,eff}}{\gamma_\mathrm{prod,eff}} \approx \left( \frac{T_{e,sep}}{T_\mathrm{e,LH}} \right)^{1.5}
    \label{eq:gamma-rates-approx-tau_LH}
\end{equation}
where the ratio $T_{e,sep}/T_\mathrm{e,LH}$ expresses the proximity to the L-H boundary. The exponent 1.5 is motivated by the fact that the rates ratio scales as $\gamma_\mathrm{suppr,eff}/ \gamma_\mathrm{prod,eff}\propto T_{e,sep}^{1.5}$ across the wider database shown in~\fref{fig:H-mode-suppr-fav-unfav}.

To this end for each temperature point up to $T_\mathrm{e,LH}(n_e)$ and the other known parameters the effective production rate $\gamma_\mathrm{prod,eff}$ is estimated following the approximations on the right-hand side of \eref{eq:H-suppr-gamma-eff}. 
However, in L-mode a different scaling for $\lambda_{p_e}$ than the one used for H-mode is required. Therefore, the L-mode scaling $\lambda_{p_e}=17.3\alpha_t^{0.298}\,\mathrm{mm}$ from~\cite{Manz_2023} will be used. Due to the large scatter of measured $\lambda_{p_e}$ relative to the scaling, the prediction will be shown also with errorband representing $\lambda_{p_e}\pm 4$~mm which is representative of the scatter.

\begin{figure}
    \centering
    \includegraphics[width=\linewidth]{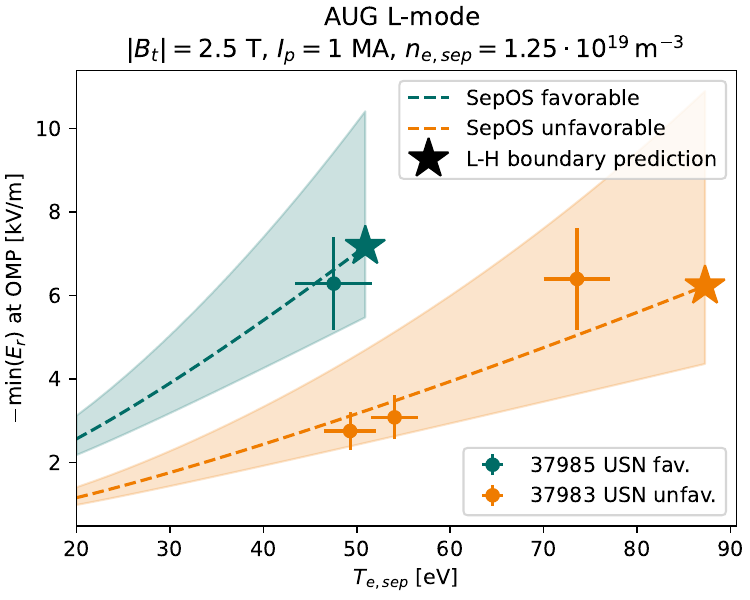}
    \caption{Predicted (negative) radial electric field field $E_r$ minimum at the outer midplane for the favorable and unfavorable configurations from the SepOS models vs. experimental measurements (circle points) reported in~\cite{Plank_Er_2023} for the medium density USN case. The star symbol shows the expected $E_r$ at the L-H boundary.}
    \label{fig:Er-min-USN}
\end{figure}

\begin{figure}
    \centering
    \includegraphics[width=\linewidth]{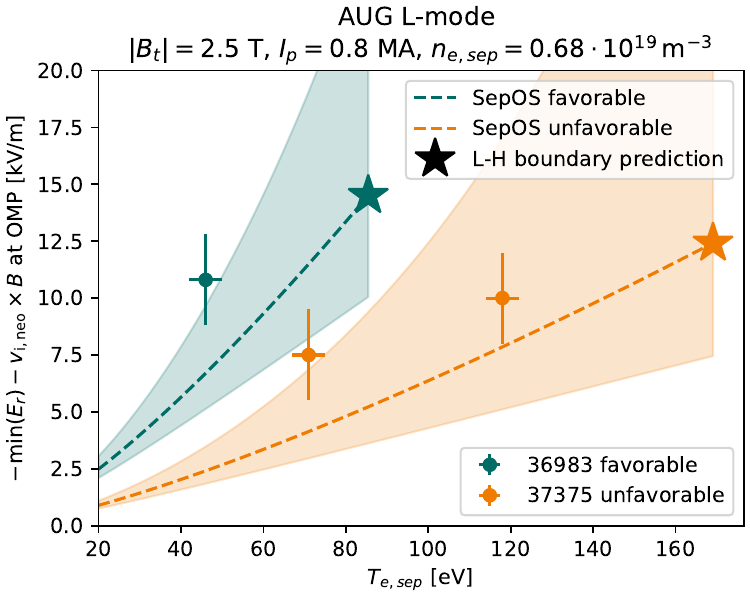}
    \caption{Predicted (negative) radial electric field field $E_r$ minimum  at the outer midplane for the favorable and unfavorable configurations from the SepOS models vs. measurements (with the toroidal component $v_\mathrm{i,neo}\times B$ estimated by NEOART  subtracted) reported in~\cite{Plank_Er_2023} for the low density LSN case. The star symbol shows the expected $E_r$ minimum at the L-H boundary.}
    \label{fig:Er-min-LSN}
\end{figure}

The combination of \eref{eq:gamma-rates-approx-tau_LH} and \eref{eq:E_r-from-suppr} gives the simulated $- \mathrm{min}(E_r^\mathrm{OMP})$ evolution from L-mode towards H-mode shown in~\fref{fig:Er-min-USN} and~\fref{fig:Er-min-LSN} as dashed lines for the favorable and unfavorable configurations. The $T_\mathrm{e,LH}(n_e)$ points in either configuration are represented by star symbols. The results of these calculations are also compared with the actual $E_r$ minimum measurements reported in~\cite{Plank_Er_2023} for the medium density, USN and low density, LSN  cases in~\fref{fig:Er-min-USN} and~\fref{fig:Er-min-LSN}, respectively. Since in the LSN cases the toroidal rotation component appears to be significant, the NEOART estimate of this component $v_{i,neo}\times B$ with little shear is subtracted and the errorbars appropriately enlarged. In the USN cases  the toroidal rotation component appears to be insignificant, therefore, the actual measured $E_r$ values are shown in~\fref{fig:Er-min-USN}. 

The measurements agree with the prediction within uncertainties quite well, given that the estimations above were expected to capture the scaling at most.  Nevertheless, it is clear that at the same heating power, i.e. the same $T_{e,sep}$, the $E_r$ minimum is predicted to be shallower in the unfavorable configuration than in the favorable one, consistent with the experimental trends.

The gradual development of $\mathrm{min}(E_r)$ dashed line  may seem at odds with experimental observations where  $\mathrm{min}(E_r)$ typically develops more abruptly as the heating power is increased towards the transition to H-mode~\cite{Plank_Er_2023}. However, it is important to note that here the figures are shown with respect to the separatrix temperature $T_{e,sep}$. If they were shown with respect to $P_\mathrm{SOL} \propto T_{e,sep}^{7/2}$, the rise would indeed appear to be much faster.

The predicted $E_r$ near the L-H boundary (star symbols) is, however, quite similar, again in agreement with the general experimental observation that the L-H transition appears to occur at roughly the same shearing rate or $E_r$ minimum (as the proxy of the shear with negligible or subtracted, insignificantly sheared toroidal rotation  consistent with observations in~\cite{Plank_Er_2023}) in either configuration.  The predicted values near the L-H boundary are close to the typically experimentally observed $E_r$ minimum~\cite{Plank_Er_2023}. The calculations would suggest that the $E_r$ near the L-H boundary is slightly lower in the unfavorable case. However, this may be due to the employed $\lambda_{p_e}$ scaling which was derived only for favorable L-mode cases which did not include  higher temperatures $>80$~eV and thus may be simply extrapolating too far outside its domain of validity.  Generally, the employed $\lambda_{p_e}$ scaling was developed to capture the general L-mode trends and did not specifically target L-modes close to the L-H transition.

The prediction of the absolute value  of $|\mathrm{min}(E_r)|$ (neglecting toroidal rotation) or equivalently $\gamma_\mathrm{suppr,eff}$  near the L-H boundary generally appears not to be entirely constant with density, possibly slightly decreasing towards higher densities. While this may be  partly due to the imperfect scaling,  it is broadly consistent with measurements reported in~\cite{Plank_Er_2023,Brida22-Er}.

Regardless of all the aforementioned uncertainties and imperfect assumptions, the main result is that the model robustly reproduces the general qualitative trend of $\mathrm{min}(E_r)$ being deeper in the favorable configuration relative to the unfavorable case at the same heating power.

\section{Discussion}

Since the criterion \ref{eq:H-suppr-conceptual} conceptually focuses on the H-L back-transition, it is worth considering how far from the L-H transition it may be. Although the L-H and H-L transitions  can be conceivably different in certain aspects, such as in the heating power hysteresis, the typical difference  is comparable to the uncertainty of the power crossing the separatrix estimate employed here. Therefore, the analysis presented herein cannot reliably distinguish the hysteresis effect anyway.  Furthermore, past studies in AUG have shown that local edge  quantities such as the pedestal top pressure (as a proxy for the flow shear driven by the pressure gradient) or the radial electric field do not exhibit the hysteresis seen in the power~\cite{Plank_2023_PPCF,Willensdorfer_2012,Wolfrum_2012}.  Therefore, as the study in this article using separatrix kinetic quantities to approximate the flow shearing of turbulence is conceptually similar, the hysteresis effects can be expected to be limited.  Finally, one might argue that for a reactor knowledge of requirements for reaching a sustainable, ``locked-in'' H-mode is of ultimate importance.

The original SepOS model \cite{Eich_2021}, includes the tilt by $k_x=k_y$. 
Structures are tilted with
 $k_x = k_y \partial_x u_y \tau$  which with the shear suppression criterion results in $k_x=k_y$.
The shear flow can interact effectively with the turbulence only if the inverse eddy turn-over time or vorticity
$\tilde{\Omega}=\nabla\times \tilde{\mathbf{u}}$
is approximately equal to the shearing rate $\partial_x u_y$. This means $\partial_x \tilde{u}_y\sim \partial_x u_y$ the radial derivatives of the binormal velocity are of similar order. Unlike for density and temperature fluctuations, the fluctuations in binormal velocity are not smaller than the background velocities, both being on the order of diamagnetic velocity. This is due to the fact that the fluctuations are small but their gradients are not,
thus
$\nabla \tilde{\phi}\approx \nabla \tilde{p}\approx \nabla p$.
As these are eddies, 
$\partial_x \tilde{u}_y\sim \partial_y \tilde{u}_x \sim k_y^2 \tilde{\phi}$. 
When the inverse eddy turn-over time is similar to the shearing rate $\partial_x \tilde{u}_y\sim \partial_x u_y$,
approximating $\langle\tilde{u}_x\tilde{u_y}\rangle \partial_x u_y$ by
$u_y \partial_x\langle \tilde{u}_x\tilde{u}_y\rangle$ 
seems reasonable. 

Nevertheless, these considerations about the fluctuation amplitude should not be interpreted as leading to a zonal flow of the same amplitude as the mean flow. This is because although the fluctuation amplitude of individual turbulent structures may be large, the average Reynolds work contribution to the total flow is still expected to be only a small fraction relative to the mean flow dominated by the mean pressure gradient.

In the SepOS model, the characteristic wavenumber is not determined by the linear growth rates, i.e. it is not approximated by the dominant mode. 
The main principle of linear growth rate determination by determining its maximum is not applied. Therefore, it is not a linear stability calculation. 
Intrinsically, small scales are first suppressed by the shear flow~\cite{Scott_2005,ManzPRL2009ZF-DW,Happel2011-TJ-II}. So one should look for the largest scale to be suppressed and that should be the transition from electrostatic to electromagnetic turbulence $k_{ES}$. If this scale is suppressed by the shear flow, all wavenumbers above it should also be suppressed, because typical turbulence spectra decay towards  higher wavenumbers, and at the same time the transfer term $\partial_x \avg{\tilde u_x \tilde u_y}\propto k^3$ is stronger. Turbulence on smaller wavenumbers is electromagnetic and then responsible for the remaining transport in the H-mode.
The growth rates enter as effective nonlinear growth rates, with only the cross phase being approximated quasi-linearly. This leads to similarities with the calculations for linear growth rates. 

The impact of the Reynolds stress in the end appears to not be described exclusively by either the theory in \cite{Fedorczak_2012} or \cite{StaeblerPoP2011}. Rather it is a combination of the best parts of both theories. Specifically, the theory of \cite{Fedorczak_2012} seems to have correctly recognized the importance of the competition between the magnetic and flow shear, but attributed the origin of the Reynolds stress asymmetry to the fluctuation envelope which does not appear to agree with the presented GRILLIX and GENE-X simulations. On the other hand, \cite{StaeblerPoP2011} recognized the impact of the local magnetic shear asymmetry, but did not recognize the impact of the flow shear and instead used an ad-hoc ansatz based on the $\nabla B$ drift.

It is worth stressing that neither of the GRILLIX or GENE-X simulations analyzed actually simulates H-mode-like conditions yet, nor are the matches with the experimental radial electric field (or potential) profiles particularly good. Therefore, it is possible that a better matched simulation of H-mode conditions might reveal a behavior not observed in the current analysis. Nevertheless, the very good match of the general equilibrium-based arguments with the simulation results across two completely different machines and for both favorable and unfavorable configurations offers hope that the conclusions regarding the average tilt asymmetry shape being dominated by the local magnetic shear asymmetry may be sufficiently general. Preliminary analysis of on-going AUG H-mode GRILLIX simulations~\cite{ZholobenkoH-mode-AUG} indeed suggests that the conclusions about the average tilt components following \eref{eq:RS-DALF-Fedorczak} hold  in that case as well.

The conclusions are expected to hold when the very edge of the plasma is in a turbulence regime dominated by drift-wave-interchange turbulence susceptible to magnetic shear effects, and the particle source is sufficiently localized relatively far away from this edge layer - in the sense it has limited impact on any turbulence velocity energy asymmetries.
Contrarily, the arguments above of course cannot account for differences in symmetric equilibria, such as in limited circular plasmas considered in the experimental comparison in~\cite{Fedorczak_2013}, where indeed only an asymmetry in the velocity distribution (possibly due to a particle source determined by the limiter location) can account for differences between favorable and unfavorable configurations. Also in double null diverted configurations the magnetic symmetry implies that any asymmetries in behavior will likely have a different origin, possibly the impact of a different $E_r$ field in the SOL due to different neutrals behavior. Such fine details related to a balance of particle sources and sinks could of course be very machine-dependent, possibly giving rise to often contradictory results observed in the community.

The presented results may appear to suggest that the difference between the favorable and unfavorable configuration, or even the L-H transition physics generally, is only due to the very edge of the plasma just inside the separatrix, in the outer shear layer at the pedestal foot. This may seem to be at odds with the observations that the shear suppression occurs first around the inner shear layer deeper inside~\cite{Cavedon-inner-shear} and more generally with the whole pedestal region build-up during the L-H transition. However, this is not in contradiction, because the conditions at the outer shear layer in the vicinity of the separatrix can be seen rather as a necessary, but  not sufficient, condition for the establishment of a fully turbulence-suppressed pedestal. The choice of this position in the vicinity of the separatrix is rather a convenient balance between several motivations: Firstly, it enables a direct connection to power exhaust considerations, which makes this reduced model particularly well-suited for reactor design considerations. Secondly, in this region the very simple drift-fluid model basis is likely still applicable and exploitable for such semi-analytical, quasi-linear estimations.

The general physical model of the Reynolds work energy transfer being different and contributing less to turbulence suppression by shear flows in the unfavorable configuration is qualitatively similar to the observations in Ref.~\cite{Cziegler2017PRL} where the energy transfer through the Reynolds stress to zonal flows in the unfavorable configuration was found to be weaker due to simultaneous transfer to Geodesic acoustic modes (GAM).  
The overall model for the turbulence suppression  in H-mode by the transfer of turbulence energy to the shear flow is conceptually similar to related models using a similar transfer term~\cite{Tynan_2013,Tynan_2016}. Some of the differences with respect to these models is the explicit quasi-linear approximation of the production terms in \eref{eq:H-suppr-gamma-eff} and the approximation of the mean flow through only the diamagnetic component in \eref{eq:u_y-general}. Specifically, in the above the generated zonal flow contribution to the mean flow is considered to be negligible relative to the diamagnetic component due to the low turbulence level in H-mode. The impact of the mean flow shear on the production terms also isn't taken directly into account. The extension of the model towards these considerations may offer opportunities to describe phenomena such as  limit cycle oscillations as was done in~\cite{Tynan_2013}.

\section{Conclusion and summary}

    The combination of the SepOS model~\cite{Eich_2021} and a combination of ideas about the magnetic-shear-induced Reynolds stress symmetry breaking from~\cite{Fedorczak_2012} and~\cite{StaeblerPoP2011} provides a model which can quantitatively match the experimental observations of the increased power threshold in the unfavorable ion $\nabla B$ drift configuration in ASDEX Upgrade. The main mechanism for increasing the power threshold is the weakening of the Reynolds stress which acts as a ``conduit'' for transferring turbulence energy to the mean flow, thereby maintaining turbulence suppression. The Reynolds stress is weakened  due to the flow shear acting on average against the local magnetic shear in unfavorable configuration. This mechanism has been quantified using the average structure tilt. The total average tilt is determined mainly by the combination of the toroidal magnetic field direction determining the mean flow shear direction and the local magnetic shear asymmetry induced by the single null X-point geometry.
    
    The quantification of the average local magnetic shear effect in terms of the average structure tilt in several GRILLIX and GENE-X simulations leads to a common estimate which agrees well with the observed favorable/unfavorable difference describing the larger ASDEX Upgrade database.
    Interestingly, the model describes the difference between the favorable and unfavorable configurations consistently with the usual $\nabla B$ drift orientation, though for physics reasons not directly related to the $\nabla B$ drift.

Additionally, the combined model can also qualitatively explain the difference in the radial electric field $E_r$  between the favorable and unfavorable configurations at the same heating power, as seen experimentally. The predicted $E_r$ minimum values appear to even quantitatively match quite well the experimentally observed values, though due to the uncertainties involved a claim of a full quantitative match of $E_r$ would be premature. 

Future efforts will focus on the comparison of this model with upcoming H-mode GRILLIX simulations, as well as with other simulation codes, in order to determine how general the picture established here is. Similarly, the establishment and analysis of a larger, multi-machine  database of separatrix operational space quantities in the framework of an on-going ITPA activity is expected to eventually yield also validation  of this model on a wider scale from the experimental side.

\section*{Acknowledgements}

The authors would like to thank U. Stroth, E. Wolfrum, M. Dunne and G. Birkenmeier for fruitful discussions and comments. This work has been carried out within the framework of the EUROfusion Consortium, funded by the European Union via the Euratom Research and Training Programme (Grant Agreement No 101052200 — EUROfusion). Views and opinions expressed are however those of the author(s) only and do not necessarily reflect those of the European Union or the European Commission. Neither the European Union nor the European Commission can be held responsible for them.

\appendix

\section{Reynolds stress sign expressed through wavenumbers}
\label{sec:RS-sign}

For the derivation of the minus sign in \eref{eq:RS-DALF} it is important to note that the Reynolds stress as the covariance of the velocity fluctuations calculated over some $(x,y)$ domain can be understood as the 0-shift cross-correlation $(\cdot\star\cdot)$ 
\begin{equation}
    \langle \tilde u_x \tilde u_y\rangle =\int\int\tilde u_x \tilde u_y \mathrm{d}x \mathrm{d}y= (\tilde u_x \star \tilde u_y)(\Delta x=0,\Delta y=0)
\end{equation}
This translates in the Fourier domain to a product of the Fourier components\footnote{strictly speaking, it is a Fourier series in $k_y$ since $y$ is periodic in the confined region, but in the interest of clear visual presentation this detail is not noted explicitly with a sum symbol.}, but with one of the complex conjugated as derived similarly for the time domain in~\cite{RS-BPP-COMPASS}
\begin{eqnarray}
     (\tilde u_x \star \tilde u_y)(\Delta& x,\Delta y) = 
     \int\int  \hat{u}_x(k_x,k_y)\cdot  \overline{\hat{u}_y(k_x,k_y)} \cdot \nonumber\\
     &\cdot \exp(ik_x \Delta x  + ik_y \Delta y) \mathrm{d}k_x \mathrm{d}k_y
\end{eqnarray}
where the hat accent $\hat u$ signifies the Fourier component and $\bar u$ complex conjugation. Therefore, for the covariance as the 0-shift cross-correlation the following holds 
\begin{eqnarray}
    \langle \tilde u_x \tilde u_y\rangle &= (\tilde u_x \star \tilde u_y)(\Delta x=0,\Delta y=0) =\nonumber\\
    &=\int\int \hat{u}_x(k_x, k_y)\cdot   \overline{\hat{u}_y(k_x,k_y)}  \mathrm{d}k_x \mathrm{d}k_y
\end{eqnarray}

Finally,  using the approximation of a single $(k_x,k_y)$ dominant mode, and approximating the velocity fluctuations as $E\times B $ velocities $\hat{u}_x \approx i k_y\hat\phi$ and $\hat {u}_y\approx -i k_x\hat\phi$, the complex conjugation plays a role
\begin{equation}
    \hat{u}_x \overline{\hat{u}_y} \approx (i k_y \hat\phi) \overline{(-i k_x\hat\phi)}=-k_x k_y \tilde\phi^2
\end{equation}
Note that the differing sign orientation of the $E\times B $ velocity components coming from the vector product definition is specific to the right-handed $x,\varphi, y$ coordinate system.  Fortunately, the end result is invariant under a change to a different (right-handed) system like $x,y, \varphi$, because the sign would flip but the product would remain the same. For completeness let us note that the Reynolds stress is invariant under a $B$ field direction change, because the product implicitly carries a square $1/B^2$.

For comparisons between articles it is of note that the minus sign in \eref{eq:RS-DALF} was also obtained in \cite{Fedorczak_2012,Fedorczak_2013}, though the phase shift represented by the imaginary unit $i$  was actually neglected and the wavenumbers were used rather as a gradient scale. However, in the more recent version of those considerations in~\cite{Peret2022} the phase shift is apparently taken into account, but not the complex conjugation, leading to an apparent lack of the minus sign. The complex conjugation was also omitted in the derivation in~\cite{Eich_2021}.

\printbibliography

\end{document}